\documentclass[fleqn,10pt]{wlscirep}
\usepackage[utf8]{inputenc}
\usepackage[T1]{fontenc}
\usepackage{xcolor}
\usepackage{colortbl}
\usepackage{amssymb}% http://ctan.org/pkg/amssymb
\usepackage{pifont}
\title{A multimodal
stress detection dataset with facial expressions and physiological signals}
\usepackage{graphicx}
\usepackage{subcaption}
\usepackage{amsmath,amssymb,amsfonts}
\usepackage{multirow}
\usepackage{algorithmic}
\usepackage{textcomp}
\usepackage{float}
\usepackage{multicol}
\usepackage{xcolor}
\usepackage{colortbl}
\usepackage{tablefootnote}
\usepackage{longtable}
\usepackage{soul}
\restylefloat{table}  

\author[1]{Majid Hosseini}
\author[2]{Fahad Sohrab}
\author[1,*]{Raju Gottumukkala}
\author[1]{Ravi Teja Bhupatiraju}
\author[1]{Satya Katragadda}
\author[3]{Jenni Raitoharju}
\author[2]{Alexandros Iosifidis}
\author[2]{Moncef Gabbouj}

\affil[1]{University of Louisiana at Lafayette, Lafayette, LA, USA}

\affil[2]{Tampere University, Tampere, Finland}
\affil[3]{University of Jyväskylä, Jyväskylä, Finland}
\affil[*]{Raju.Gottumukkala@louisiana.edu}


\begin{abstract}
Affective computing has garnered the attention and interest of researchers in recent years, as there is a need for AI systems to better understand and react to human emotions. However, analyzing human emotions, such as mood or stress, is quite complex. While various stress studies use facial expressions and wearables, most existing datasets rely on processing data from a single modality. This paper presents EmpathicSchool, a novel dataset that captures facial expressions and the associated physiological signals, such as heart rate, electrodermal activity, and skin temperature, under different stress levels. The data was collected from 30 participants during different sessions for about ninety minutes each (for a total of 40 hours). The data includes seven different signal types, including both computer vision and physiological features that can be used to detect stress. In addition, various experiments were conducted to validate the signal quality.

\end{abstract}
\begin{document}

\flushbottom
\maketitle
\thispagestyle{empty}

\section*{Background and summary}
Affective computing is an interdisciplinary field aimed at developing systems that can identify, recognize, and interpret human emotions  \cite{poria2017review}. Affective computing leverages physiological signals to detect the individual's emotions or state of stress. Stress is one of the major problems in modern society that leads to significant health expenditure \cite{can2019stress}. Early ambulatory stress detection has become an exciting challenge \cite{plarre2011continuous,smets2018large} because early stress detection can mitigate the conversion of stress to a chronic disease \cite{can2019stress} and prevent stress-related health issues \cite{wijsman2011towards}. Human emotions manifest through physiological, behavioral, and cognitive changes, which can be utilized to adapt digital environments. For example, the current state of a smartphone user can be continuously computed based on their interactions with various digital services \cite{karikoski2013contextual}, allowing personalized services to be provided in response to the user's mental state  \cite{politou2017survey}.  Additionally, stress factors, such as high workload, lack of autonomy, and long working hours, can negatively impact people's health \cite{wilson2009fear}. Many studies point out that prolonged exposure to stress leads to chronic conditions, such as obesity  \cite{peternel2012presence} or hypertension  \cite{bickford2005stress}, which may exacerbate conditions, such as type-II diabetes  \cite{wellen2005inflammation}. Therefore, monitoring and understanding stress in workplaces is essential, especially in professions with increased exposure to stress, often leading to burnout and increased turnover  \cite{greenglass2001workload}. 

Stress is primarily a physiological response to a stimulus, typically triggered by an external factor in an environment \cite{chu2024physiology,filipovic2001review}. Epinephrine, commonly known as adrenaline, is a significant stress hormone \cite{hoffman2013adrenaline}. In a daily routine, a moderate amount of stress is considered beneficial \cite{awada2024stress}. It causes gentle excitement and improves performance when carrying out regular daily tasks. A reasonable amount of stress that causes excitement is termed eustress or positive stress \cite{li2016eustress,qian2014does}.
On the other hand, undesirable stress is called distress or negative stress \cite{qian2014does,hill2018sense}. Repeated stress events can lead to emotional exhaustion \cite{jin2020impact}, and the resulting long-term physiological stress can be harmful; such stress is termed chronic stress \cite{mcewen2006protective}. The effects of chronic stress may include anxiety, depression, sleep disturbances, and weight gain \cite{schneiderman2005stress}. It may even compromise immune responses, increasing vulnerability to several diseases  \cite{maslach1998multidimensional}. Studies related to stress have gained interest in recent years due to its widespread effects on health, family, workplace productivity, society, and the economy \cite{murphy2003usa}.

The analysis of stress involves using wearable sensors to deduce physiological signals. Table \ref{tab:signal} shows the standard signals provided by wearables. In contrast, the study and analysis of human emotions use a camera to infer facial expressions that represent emotions. In the literature, emotions and stress signals are typically considered two disjoint topics and are usually studied independently. Schmidt et al. \cite{schmidt2018introducing} identify the gap in analyzing basic emotions and stress together; however, there is still a need for a dataset that addresses the issue of correlating stress and facial expressions under different stressful situations. This work presents a multimodal stress-emotion dataset containing stress data from wearable devices and emotion data extracted from facial expressions through a video feed. We analyze human expressions and physiological signals under different stress levels. The dataset presented is the first effort toward identifying stress from non-wearable devices in general and facial expressions in particular. The current study is inspired by our previous work on stress detection of nurses using wearable devices  \cite{hosseini2021multi} and our initial effort to use facial expressions to study satisfaction  \cite{sohrab2020facial}. This study collected stress and facial expression data from 30 participants under different stress levels. 

\begin{table}[ht]
\caption{Common physiological signals.}
\begin{center}
\definecolor{Gray}{gray}{0.85}
\definecolor{LC}{rgb}{0.7,.9,.9}
\definecolor{LB}{rgb}{0.9,.9,1}

\begin{tabular}{ | m{3.8cm}| m{5.5cm} | m{.9cm}|}
\hline
   \textbf {Signal}   &\textbf{Device}                        &\textbf{Abbr}\\            \hline
    Heart Activity                  &Electrocardiography                       &ECG EKG\\                  \hline
    Skin Response      &Electrodermal Activity                 & EDA\\                     \hline
    Skin Response                   & Galvanic Skin Response                &GSR\\                      \hline
    Muscle Activity    &Electromyography                       &EMG\\                      \hline
    Respiratory Response            &Electromagnetic Generation             &RR\\                       \hline
    Respiratory Response&Respiratory Inductive Plethysmograph  &RIP\\                      \hline
    Blood Volume Pulse              & Cardiovascular Dynamics               &BVP HR HRV\\               \hline
    Skin Temperature   &Thermistor                             &TEMP\\                       \hline
    Brain Activity                  &Electroencephalogram                   &EEG\\                      \hline
    Eye Activity       &Corneo-retinal Standing                &EOG\\                      \hline
    Physical Activity               &3-axis Accelerometer                   &ACC\\                      \hline
    Physical Activity  &3-axis Magnetometer                    &MGM\\                      \hline
    Heart Activity                  &Sphygmomanometer                       &BP\\                       \hline
    Eye Characteristics&Pupil Diameter Detector Module         &PD\\                       \hline
    Body Temperature                & Thermal Imaging                       &TI\\                       \hline

\end{tabular}
\label{tab:signal}
\end{center}
\end{table}

There is a host of surveys in the stress research literature. Alberdi et al. \cite{sioni2015stress} listed physiological signals used for stress detection. Fukazawa et al. \cite{thapliyal2017stress} introduced standard stress detection tools and devices. Carneiro \cite{carneiro2017new} described techniques to access and monitor stress in offices and other workplaces. Alberdi et al. \cite{alberdi2016towards} provided a comprehensive survey on stress recognition for office environments. They described the signals and their related features, as well as successful neural network models that detect stress. Using wearables and smartphones, \cite{can2019stress} investigated stress in daily life. They briefly described stress and its impact on society and investigated the stress tests and related signals. However, they did not discuss instruments and devices detecting stress, signal properties, machine learning features, or training parameters. In another paper, Fukazawa et al. \cite{fukazawa2019predicting} provided a comprehensive survey of stress detection signals and techniques from a machine learning perspective. They only used location, activity, phone usage, context, sleep, and speech features to detect stress. They covered several papers that used a specific signal for stress detection and investigated how commonly a method or signal is used by showing its using percentage across the years. Panicker et al. \cite{panicker2019survey} presented a comprehensive survey on stress detection. They provided a comprehensive description of emotions and their organismic subsystems. They also provided definitions of stress, enumerated stress detection signals and devices, presented three different stages of stress, and evaluated the correlation and differences between different types of stress and emotions. 
Zhang et al. \cite{zhang2020emotion} explored the integration of multiple physiological signals, including EEG, EMG, GSR, and ECG, for emotion detection across various daily activities. By constructing a deep neural network architecture, they achieved 95 percent accuracy in binary classification of arousal by combining task-specific representations for each signal type. The proposed LSTM-based model demonstrated high accuracy, emphasizing the complementary nature of physiological and visual stress indicators. Kurniawan \cite{kurniawan2013stress} investigated several classification methods, such as SVM and Gaussian Mixture Models (GMM), to detect stress levels using GSR and speech features. They gained between 70 to 80 percent accuracy using GSR and 92 percent accuracy by combining both the speech and GSR using four classification techniques, namely K-means, decision tree, GMM, and SVM classification models. Moreover, some authors use thermal images to detect stress or deception detection \cite{pavlidis2005system}.
 
The AffectiveRoad dataset \cite{haouij2018affectiveroad} contains physiological signals to study the stress level of 10 drivers while driving in various environments. The study was conducted with drivers taking an 86-minute driving test in Tunisia. Schmidt et al. \cite{schmidt2018introducing} introduced the WESAD (Wearable Stress and Affect Detection) dataset and studied the stress of 15 students while watching a movie and taking a TSST (Trier Social Stress Test) test \cite{kirschbaum1993trier} using E4 signals. MDPSD (multimodal dataset for psychological stress detection) \cite{chen2021introducing} is a multimodal stress detection dataset of EDA and PPG signals collected from university students while performing different tests (e.g., TSST \cite{birkett2011trier}, IQ test, and color-word tests \cite{scarpina2017stroop}). Mundnich et al.\cite{mundnich2020tiles} provided TILES-2018, a multi-sensor dataset that provides a battery of surveys to cover personality traits, behavioral states, job performance, and well-being over time. The SWELL \cite{koldijk2014swell} dataset provides data corresponding to different stress levels of participants while performing some office work (e.g., answering email) at three different stress levels (neutral, stressor, and stressor with interruption) using Kinect, ECG, and emotional expressions extracted from videos. However, this dataset does not provide the videos of the participants and uses outdated facial features to detect emotion and stress. In addition, the SWELL dataset provides the stress level of the entire task as one label. Zaman et al. \cite{zaman2019stress} studied the impact of email interruptions and office activities on productivity and stress, using multimodal data from 63 workers who were exposed to batch or continual email interruptions, with or without additional stress. The data collected, including physiological stress indicators, writing quality measures, keystroke dynamics, and participant profiles, is expected to provide insights into personalized email management strategies and a better understanding of office activity dynamics.

Although WESAD \cite{schmidt2018introducing} and SWELL \cite{koldijk2014swell} have been invaluable for stress detection research, they have some limitations. WESAD provides synchronized ECG, EDA, and motion data, but no facial video or expression labels. This makes WESAD single‐modal from the vision standpoint and unable to capture behavioral cues. SWELL adds facial‐expression features, but only reports them as coarse, single‐label summaries per task (neutral vs. stressed), does not provide the raw video data, and relies on feature extractors, preventing reanalysis or application of new deep learning methods. Furthermore, both datasets offer workload labels at session-level granularity, which limits analysis of rapid stress fluctuations. As a result, there is a need for a multimodal, high‐resolution, open dataset that couples raw facial video and physiological streams with minute‐level stress annotations.

Our dataset, EmpathicSchool, provides physiological stress signals streams collected from Empatica E4 and facial features from videos of engineering and computer science students performing different tasks. Our primary motivation for creating this dataset was to generate a multimodal emotion-stress detection dataset and analyze students' emotions and stress levels while performing routine tasks such as preparing for a presentation or taking an exam. In addition, we analyze the emotions and stress levels of the students after task fulfillment. To summarize, EmpathicSchool addresses the gaps observed in the existing datasets in four principal ways:
\begin{itemize}
  \item \textbf{True multimodality}: we release both raw facial‐video 1080p at 30 fps frame rate and synchronized physiological data (EDA, BVP, ACC, HR, and TEMP), enabling joint vision-biometric signal modeling.
  \item \textbf{Fine‐grained labels}: self‐assessment stress and NASA-TLX workload scores are provided at 2-minute intervals, allowing detection of transient stress changes, unlike the task‐level single labels in SWELL.
  \item \textbf{High temporal resolution}: all modalities are timestamp-aligned, and video frames and biometric samples can be precisely upsampled or downsampled to any common rate up to 30 Hz, supporting both frame-based and windowed analysis.
  \item \textbf{Open, extensible data}: we publish raw videos, landmark files, raw E4 streams, and all code for feature extraction and preprocessing so new facial‐expression or biometric signal feature extraction algorithms can be applied directly.
\end{itemize}
By combining these improvements, EmpathicSchool substantially broadens the scope for research in near-real-time, multimodal stress detection.

\section*{Methods}
This section briefly introduces the sensors and protocols used in the data collection process. We also discuss the methods used to extract features from the video data and provide a summary of the sensors used for biomedical data acquisition.

The data was collected in two locations: Tampere University, Finland, and the University of Louisiana at Lafayette, USA. We received approval from the institutional review boards of the University of Louisiana at Lafayette (SP-21-144-IRI) and the Tampere region's ethics committee (50/2021) for the study's protocol for the corresponding locations. Before data collection, each participant received an IRB-approved presentation detailing the nature and duration of the experimental tasks, the exact signals to be captured by the Empatica E4 wristband (including electrodermal activity, blood-volume pulse, skin temperature, and tri-axial acceleration) and a laptop camera, and the intention to publish de-identified physiological time-series with either their videos or facial-landmark data in an open repository, with a suggested consent revocation window of two months after data collection. For the data collected at Tampere University, participants may request the withdrawal of their data at any time prior to the publication of the paper. After publication, withdrawal is no longer possible, as all personally identifiable information is permanently removed and the data is fully anonymized. A written informed consent was then obtained in two stages. The main consent form authorized data collection and explicitly granted permission for public dissemination of the de-identified dataset, while preserving the participant’s right to withdraw at any point before release. Immediately after the experiment, a separate debriefing form required participants to choose whether they would (a) share the de-identified facial-feature data only or (b) share both the video recordings and facial-feature data; this second document was also signed by all the participants in this dataset. All personal identifiers are stored separately under lock and key (consent forms) and on an encrypted server (video data) for six months after the data collection and before destruction, ensuring compliance with the confidentiality safeguards mandated by both ethics committees. No waiver of consent was requested was requested from the ethical review board.

\subsection*{Participants}
The participants were recruited via flyers and recruitment emails in the engineering and computer science departments of the University of Louisiana at Lafayette and via email at Tampere University.
We recruited 31 students, and after the data collection, 1 subject asked to be removed from the dataset. As a result, we had data from 30 participants aged 21 to 35 (Mean=25.3, Standard-Deviation=4.3). We asked participants (S7-S30) in Louisiana not to drink alcohol or coffee 24 hours before data collection. No such restrictions were applied to participants S1-S6 in Tampere. The exclusion criteria were pregnancy, heavy smoking, mental disorders, and chronic or cardiovascular diseases.  Data collection in the United States was conducted in two phases to achieve the suggested minimum of 30 subjects for studies involving physiological and behavioral data \cite{vanvoorhis2007understanding}.

\subsection*{Video data}
\begin{figure}[htbp]
    \centering
    \includegraphics[width = 0.3 \textwidth]{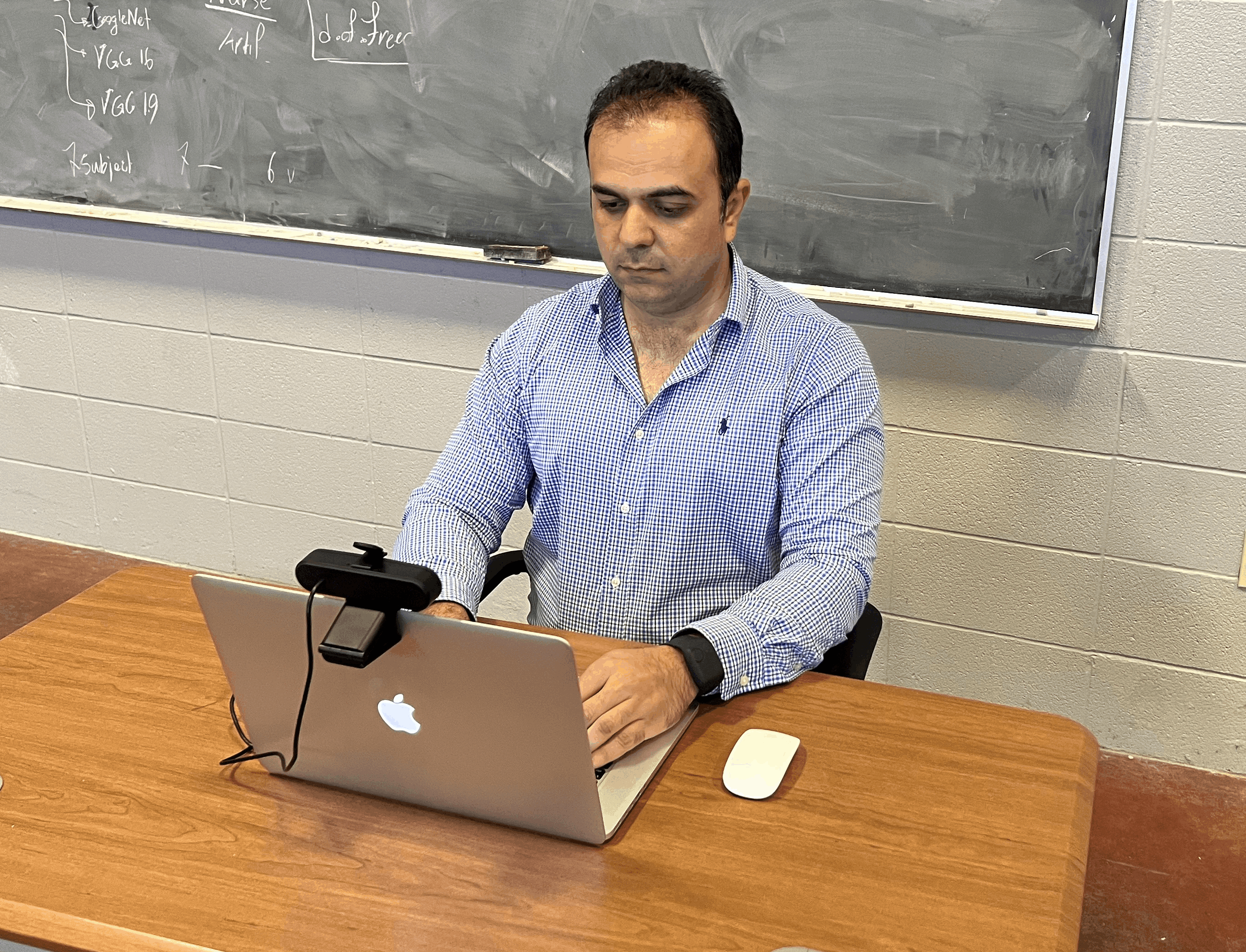}
    \caption{Participants' sitting and camera positions.  *This image is captured from one of the authors who consented to have his image in the paper; we are not providing any of the participants' images in the manuscript.}
    \label{fig:aparat}
\end{figure}

We asked the participants to perform all the tasks in a sitting position, and the camera captured only their faces and shoulders. The videos of the participants were collected using a 1080p external webcam mounted on top of a laptop screen with 30 frames per second. As approved by the institutional review board at the University of Louisiana at Lafayette, the video data of the participants who consented to release their video files are provided as part of the dataset. We post-processed the recorded videos to extract facial expressions as discussed in the following subsections. Figure \ref{fig:aparat} illustrates the experimental setup and the camera's position. 

\subsubsection*{Facial expression recognition}
\begin{figure}[htbp]
    \centering
    \includegraphics[width = 0.4 \textwidth]{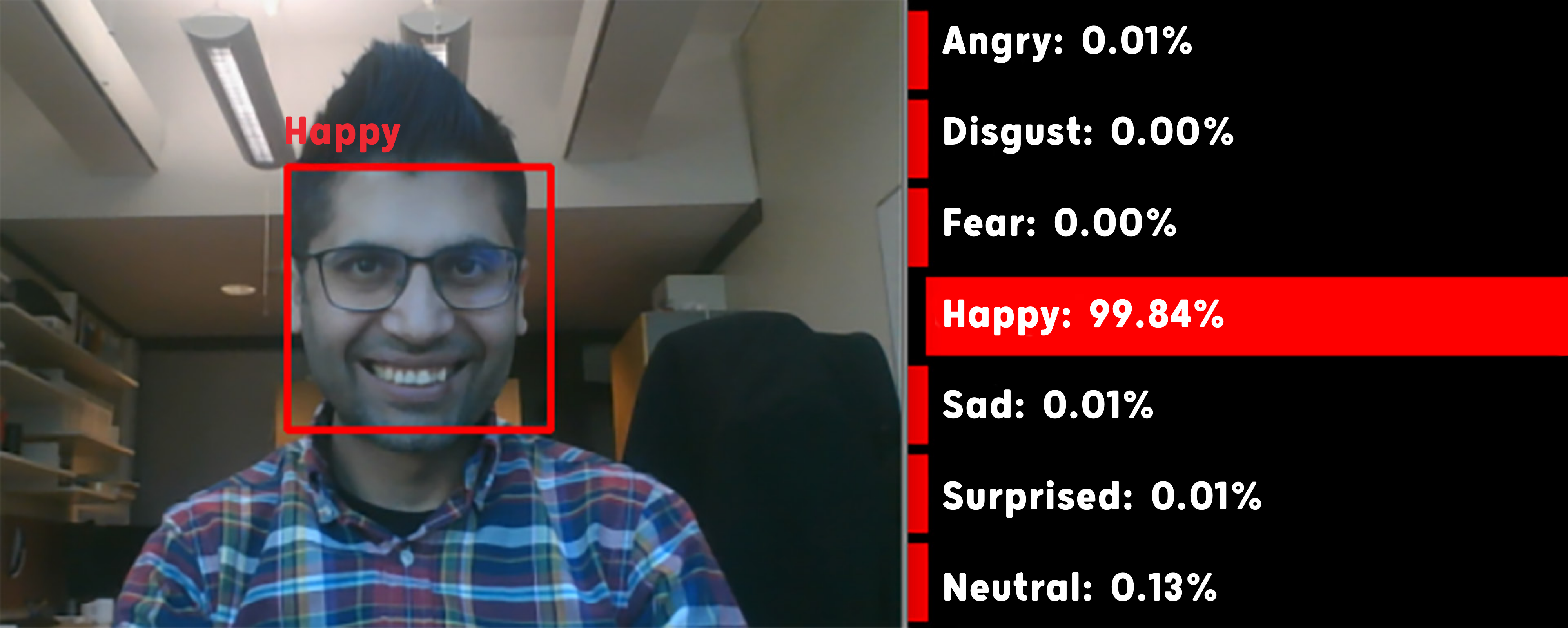}
    \caption{The deployed system captures and displays emotions from live video frames, using the Haar cascade algorithm for face detection and the MiniXception model for facial expression recognition. *This image is captured from one of the authors who consented to have his image in the paper; we are not providing any of the participants' images in the manuscript.}
    \label{fig:emotions}
\end{figure}

The facial expression recognition system consists of two modules. In the first module, we used the Haar cascade \cite{viola2001rapid} frontal face detection algorithm implemented in the OpenCV \cite{bradski2000opencv} library to detect the face in the video frame. Haar cascade uses Haar-like features to encode the local appearance of faces \cite{padilla2012evaluation}. The detected faces were processed and passed through a pre-trained model to recognize facial expressions in the second module. We used MiniXception \cite{arriaga2017real}, trained over the Facial Expression Recognition 2013 (FER-2013) dataset \cite{goodfellow2013challenges} for facial expression recognition. FER-2013 dataset contains 28,709 training images, 3,589 validation images, and 3,589 test images from the seven basic facial expression categories (0=Angry, 1=Disgust, 2=Fear, 3=Happy, 4=Sad, 5=Surprise, 6=Neutral). Our experiments deduced the facial expressions from all the video recordings. Figure \ref{fig:emotions} demonstrates real-time emotion display through the utilization of the Haar cascade algorithm for face detection in conjunction with the MiniXception model for facial expression recognition. These data are provided as part of the dataset.

\subsubsection*{Dlib features}
\begin{figure}[htbp]
    \centering
    \includegraphics[width=0.5\linewidth]{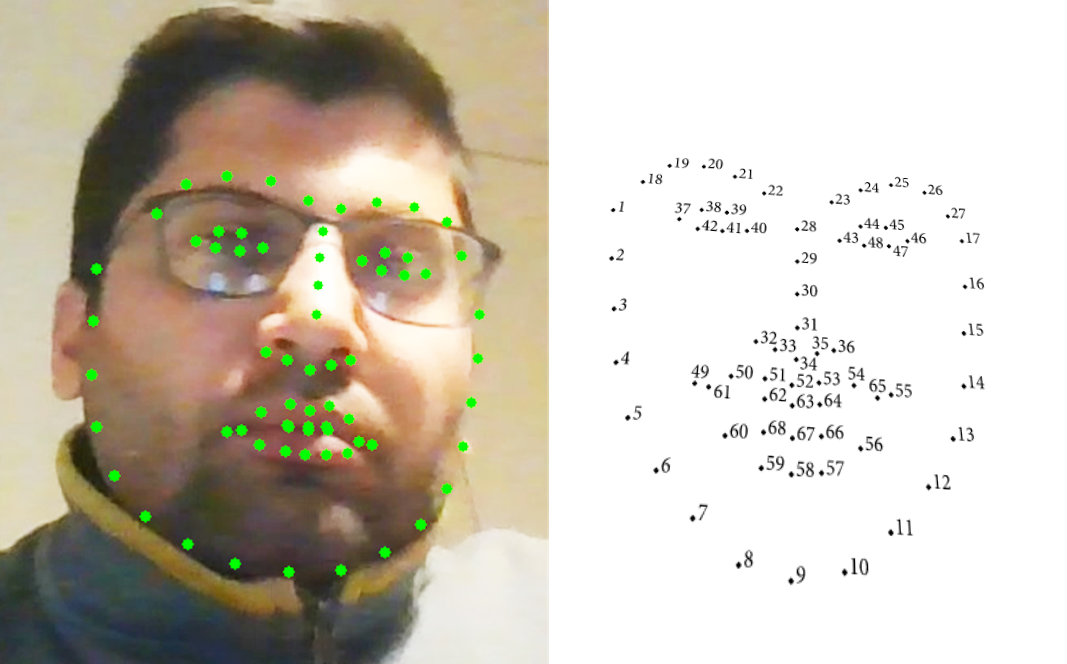}
    \caption{68 facial landmarks. *This image is captured from one of the authors who consented to have his image in the paper; we are not providing any of the participants' images in the manuscript.}
    \label{fig:68lndmrks}
\end{figure}
Facial landmarks are used in various computer vision tasks, such as drowsiness detection \cite{mohanty2019design}, fatigue detection \cite{irtija2018fatigue}, facial expression recognition  \cite{heidari202progressive}, and micro-expression classification \cite{kumar2021micro}. Therefore, we provide the facial landmarks of all the video frames collected as part of the dataset. We utilize Dlib \cite{king2009dlib}, which returns the position of 68 facial landmarks. An example of landmark detection on a facial image based on Dlib is shown in Figure \ref{fig:68lndmrks}.

\subsubsection*{Facial Action Units}

Facial Action Coding System (FACS) is a comprehensive framework for understanding facial expressions, particularly the discrete movements and gestures referred to as Action Units (AUs). Recent studies have leveraged the granularity of the FACS to analyze stress levels through facial expressions \cite{giannakakis2020automatic, PredictingDASS2019, AutomaticStressAU2020, AutomaticStressFACS2021 }.
We do not provide pre-computed AUs in the dataset. However, we provide the code to generate Facial Action Unit features from facial landmarks features in our GitHub repository \cite{Majid2022}.

\subsection*{Physiological signals/Biometric data}
We used the E4 wristband (Empatica Inc., Milano, Italy) for physiological data acquisition. E4 is a medical-grade wearable device that offers real-time acquisition of EDA, heart rate, skin temperature, and accelerometer data from the subject's dominant hand. EDA is measured via E4's silver (Ag) electrode (valid range [0.01-100] $\mu$S), while heart rate is measured via E4's photoplethysmographic (PPG) sensor. Using Bluetooth, E4 transmits data to the subject's smartphone. All data collected from the E4 wristband and the sampling frequencies are presented in Table~\ref{tab:E4Signals}.

    \begin{table}[htbp]
    \centering
    \caption{Signals and the sampling frequency of Empatica E4 sensors}
    \label{tab:E4Signals}
    
    \begin{tabular}{|l|l|l|}
    \hline
     \textbf{Signal} & \textbf{Abbreviation} & \textbf{Frequency} \\ 
    \hline
    Electrodermal Activity & EDA & 4.0 Hz \\
     Heart Rate & HR & 1.0 Hz \\
    Skin Temperature & ST & 1.0 Hz \\
     Accelerometer & ACC & 32 Hz \\
    Inter-Beat Interval & IBI & ---- \\ 
    
     Blood Volume Pulse & BVP & 64 Hz\\
    \hline
    \end{tabular}
    
    \end{table}
    
The physiological signals that we measured via the E4 wristband and are provided in our dataset are as follows:
\begin{itemize}
         
\item{\textbf{Heart rate:}}
The heart rate of a healthy individual, irrespective of gender, ranges from 60 to 100 beats per minute at rest \cite{fox2007resting}. However, the heart rate varies significantly with activity or emotional state \cite{shi2017differences}. A high heart rate is generally associated with stressful situations \cite{taelman2009influence, Kim2018StressHRV}, but a high heart rate should not by itself be interpreted as high stress \cite{schubert2009effects}. Heart rate is not primary data and is generated from the BVP signal by the Empatica E4 device.
 \item{\textbf{Skin temperature:}}
Skin temperature varies for various activities due to skin blood flow \cite{tanda2021simplified}. Skin temperature ranges from 33.5 to 36.93 Celsius degrees \cite{bierman1936temperature}. However, this can vary quite widely based on the type and length of activity and ambient temperature \cite{choi2013skin}. 

\item{\textbf{Electrodermal activity:}}
   EDA measures the amount of sweat gland activity by calculating the skin's electrical conductance using silver-chloride electrodes. The EDA signal is measured in units of micro-siemens ($\mu S$). Stadler et al, \cite{stadler2018electrodermal} mention that EDA peaks are event-related and can be a good estimator of the body's response to the stimulus \cite{Boucsein2012Electrodermal}. Collecting EDA signal using wristbands is not very accurate, and some artifacts can affect the data collection results. Additionally, in some subjects, the EDA signal cannot be captured, or it can be inaccurate. However, the purpose of this study is to use the data in multimodal stress detection, which remains robust in the absence of a modality or noisy data. We believe that the combination of video and physiological signals can help researchers detect stress using off-the-shelf wristbands and video, despite imperfections in any single signal stream.

 \item{\textbf{Accelerometer data:}} 
Accelerometer sensors can be used for multiple tasks (e.g., human action recognition and step counting) \cite{bayat2014study}. By measuring orientation and acceleration force, the accelerometer sensor can determine the device's orientation (horizontal or vertical) and the type of movement \cite{cudejko2022validity}. The accelerometers vary in type (digital and analog), sensitivity, and number of axes. Our dataset provides three-axis accelerometer data that measures the orientation of the sensor in three dimensions, which enables capturing the activity more precisely.

\item{\textbf{Blood Volume Pulse:}}
Blood volume pulse is measured using a PPG that measures light absorption to find pulse volume peaks \cite{abay2018photoplethysmography}. Heart rate is calculated by analyzing the time intervals between consecutive peaks. Thus, the BVP signal provides valuable information about heart activity.

\item{\textbf{Inter Beat Interval:}}
The inter-beat interval is the time difference between two consecutive peaks in seconds, which can provide information about heart activity, such as heart rate. 
IBI is not primary data and is generated from the BVP signal by the Empatica E4 device.
  
\end{itemize}

\subsection*{Data collection protocol}

Our stress detection study was conducted in a laboratory setting. We aimed to detect students' stress during their typical activities in a laboratory setting, while sitting behind a computer wearing a wristband. We collected data from 30 participants under different stress levels (normal and high stress). The normal stress level was assumed when the participants were at rest or performing a task that did not require significant mental effort. For the stressful tasks, we asked participants to accomplish the given assignment in a limited time \cite{koldijk2014swell}. Each of these tasks was supposed to be performed for 10 minutes. However, 30 seconds of data was trimmed from the beginning and end of each task to synchronize the video and physiological signals. This gives us 9 minutes of data for each participant and each task. The actions for nine different data collection sessions (tasks) and the duration of each session are provided in Table \ref{task}. All subjects used the complete allotted time for all the tasks, except eight out of the 30 participants who completed delivering the presentation task (T3) earlier than the allotted time. Table \ref{task} gives the expected stress level for each task.

\begin{table}[htbp]
    \caption{List of tasks performed by participants, total duration of the data included in the release, and expected stress level for each task.}
    \label{task}
    \centering
    \definecolor{Gray}{gray}{0.85}
\definecolor{LC}{rgb}{0.7,.9,.9}
\definecolor{LB}{rgb}{0.9,.9,1}
\begin{tabular}{|cl|l|l|}

\hline
\multicolumn{1}{|l|}{\textbf{Task}} & \textbf{Action performed}                                                            & \multicolumn{1}{c|}{\textbf{Duration (min)}} & \multicolumn{1}{l|}{\textbf{Expected Stress Level}}\\ \hline
 \multicolumn{1}{|c|}{T1}                                    & Reading a magazine                                                  & 9  & \multicolumn{1}{l|}{Normal}                          \\ \hline
\multicolumn{1}{|c|}{T2}                                    & Preparing a presentation & 9  & \multicolumn{1}{l|}{Stressed}                          \\ \hline
\multicolumn{1}{|c|}{T3}                                    & Delivering the presentation                                                                         & 5-9  & \multicolumn{1}{l|}{Stressed}                        \\ \hline
\multicolumn{1}{|c|}{T4}                                    & Rest and recovery                                                     & 9  & \multicolumn{1}{l|}{Normal}                            \\ \hline
\multicolumn{1}{|c|}{T5}                                    & IQ test / Stroop Color-Word Test                                                                            & 9  & \multicolumn{1}{l|}{Stressed}                          \\ \hline
\multicolumn{1}{|c|}{T6}                                    & Listening to calm music                                       & 9 & \multicolumn{1}{l|}{Normal}                             \\ \hline
\multicolumn{1}{|c|}{T7}                                   & Watching amusing video(s)                                                                   & 9  & \multicolumn{1}{l|}{Amused}                            \\ \hline

\multicolumn{1}{|c|}{T8}                   &Controlled breathing exercise                                               & 9  & \multicolumn{1}{l|}{Normal}                                  \\ \hline
\multicolumn{1}{|c|}{T9}                   & Recovery                                               & 9  & \multicolumn{1}{l|}{Normal}                                  \\ \hline

\end{tabular} 
\end{table}

\begin{figure}[ht]
  \centering
  % rotate 90° counter-clockwise
  \includegraphics[width=.5\linewidth,origin=c]{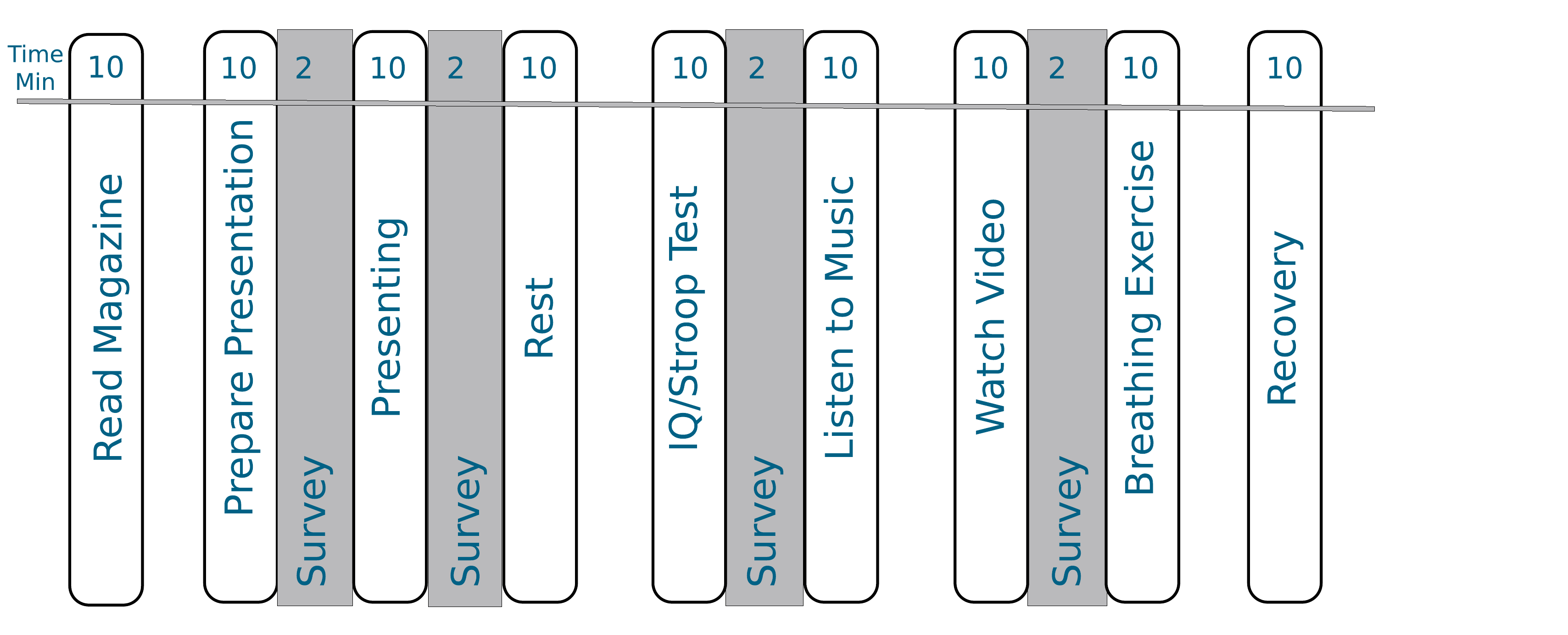}
  \caption{Schematic overview of the EmpathicSchool data collection protocol, showing the sequence and duration of each experimental phase. We do not release the data from the gray sessions.}
  \label{fig:enter-label}
\end{figure}

In task T1, the students in the United States were asked to find and read an arbitrary article from Ars Technica or Wired Internet magazines, while at the Finland site, the students were asked to read an arbitrary magazine from the Internet according to their choice to reduce their stress level to achieve a baseline level of stress \cite{knight2001relaxing}. In task T2, the participants were asked to prepare for a presentation in a limited time, followed by the presentation task T3 in the allocated time. We considered both tasks (preparing and delivering presentations) as stressful tasks. In the next task (T4), participants were asked to take a rest. In task T5, the Stroop Color-Word (Finland) or IQ test (United States) was taken by each participant to observe their stress level while taking tests in a limited time period. In task T6, we asked the participants to listen to calm music, and we considered them not to be stressed during this activity \cite{ferrer2014playing}. The rest of the data collection included watching an amusing video (T7) followed by one or two different rest activities, namely controlled breathing exercises (T8) and a recovery period (T9). Only six (S1-S6) out of 30 participants were asked to have a recovery task (T9) after a controlled breathing exercise. The video and wristband data have different start and end times, so we trimmed 60 seconds of each task for data length consistency after time synchronization (30 seconds from the beginning and the end). In our study, there is a potential for the order effect. We could not mitigate it by testing different task orders, as we lacked a subject pool of sufficient size to have statistical significance. However, we sought to minimize the order effect by introducing rest periods between the tasks as illustrated in Figure~\ref{fig:enter-label}.

Our protocol includes three stressful tasks that induce cognitive load and social-evaluative threat, identified as two dimensions of acute stress in the stress-induction literature \cite{kirschbaum1993trier,allen2017trier}. The participants prepared for a presentation under strict time pressure (T2), a manipulation that elicits moderate cognitive load and mirrors the preparation phase of the Trier Social Stress Test (TSST) \cite{kirschbaum1993trier}. They then delivered this talk to a camera while answering questions (T3), recreating the speech-and-arithmetic blocks of the TSST and, thereby, adding a social-evaluative component \cite{kirschbaum1993trier,Allen2017tsstReview}. Pure cognitive interference was induced in a timed Stroop/IQ task (T5), which elevates heart-rate and electrodermal activity \cite{scarpina2017stroop,callinan1996stroopStress}. To bracket these stressors with low-arousal or down-regulation conditions, we included quiet reading (T1) as a resting baseline commonly used in stress laboratories \cite{knight2001relaxing}, listening to calm music (T6) to induce relaxation \cite{ferrer2014playing}, and paced breathing (T8) to recruit parasympathetic activity and vagal tone \cite{laborde2021breathing}. Collectively, this sequence reproduces the most widely used laboratory stress paradigms, such as public speaking, time-pressured mental arithmetic, and Stroop interference, and embeds them between three validated recovery tasks, enabling inter-subject comparisons across the full arousal continuum \cite{kirschbaum1993trier,scarpina2017stroop,pasatReview2018}. Moreover, these tasks were related to the students' daily routines, which we believe to have the potential to facilitate research on the stress of students while performing daily tasks.

\begin{figure}[htbp]
    \centering
    \includegraphics[width=.5\textwidth]{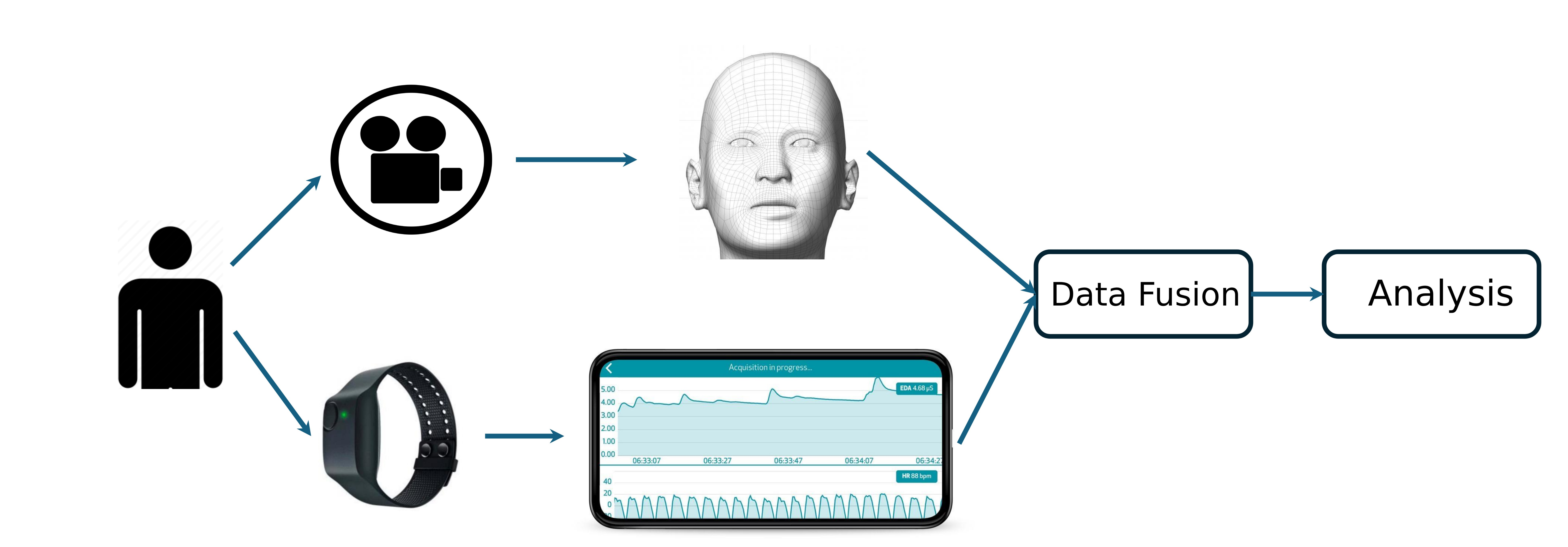}
    \caption{Data collection apparatus}
    \label{fig:apparatus}
\end{figure}

An Empatica E4 was worn on the wrist of the dominant arm, and a camera collected facial videos during the studies. We continuously collected the students' physiological data and facial features during the sessions. We detected stress events while monitoring the physiological signal streams of the students. Further, We asked participants S7-S30 to validate their stress level during each stressful task by completing a self-assessment stress level check for each of the 2-minute intervals, along with a NASA questionnaire (NASA-TLX can be further used as a proxy for stress along with the stress assessment forms). The participants were also asked to complete a questionnaire if they experienced stress during the normal stress tasks. Figure \ref{fig:apparatus} presents the framework for the data collection. A detailed explanation of all collected data is given in the Data records Section.

\subsection*{Stress detection surveys}

Stress levels can be determined via subjects' self-assessment and National Aeronautics and Space Administration Task Load Index (NASA-TLX) questionnaires \cite{hart2006nasa, hart1988development}. In this study, we provided NASA-TLX forms to participants S7-S30 and asked them to complete the questionnaires after each stressful task. Each task was divided into 5 two-minute sub-intervals, except for the first and last sub-intervals that span over 90 seconds (1.5 minutes), adding up to 9 minutes for each task. Therefore, participants (S7-S30) were asked to determine their overall stress level at each sub-interval during stressful tasks. In addition, we also asked participants (S7-S30) to rate their overall workload during different stressful tasks and provided this information and labels in the dataset. Table \ref{tab:Nasa} shows the NASA questionnaires that were asked to evaluate participants' stress levels from 0 to 20 for each sub-interval. To ensure the reliability of these scales, this study calculated Cronbach's $\alpha$ \cite{tavakol2011making} for the questionnaire scores (0.922), which shows good internal consistency. 

\begin{table}[htbp]
\centering

\caption{Workload questionnaire to evaluate the overall stress level of students during different tasks.}
\begin{tabular}{|c|l|l|}
\hline
\multicolumn{1}{|l|}{\textbf{No}} &\textbf{Category}& \textbf{Question}                                                     \\ \hline
\textbf{1}&     Mental Demand                    & How mentally demanding was the task?                                      \\ \hline
\textbf{2}&     Physical Demand                  & How physically demanding was the task?                                   \\ \hline
\textbf{3}&     Temporal Demand                  & How hurried or rushed was the pace of the task?                          \\ \hline
\textbf{4}&     Performance                      & How successful were you in accomplishing what you were asked to do?      \\ \hline
\textbf{5}&     Effort                           & How hard did you have to work to accomplish your level of performance?   \\ \hline
\textbf{6}&     Frustration                      & How insecure, discouraged, irritated, stressed, and annoyed were you?    \\ \hline

\end{tabular}
\label{tab:Nasa}
\end{table}

\section*{Data records} 

 The dataset is anonymized and available on Zenodo \cite{ZenodoDataset}. The dataset contains 30 folders (S1 to S30), one for each participant. Each folder consists of sub-folders (T1 to T9) according to the tasks performed by participants; the participants S7 to S30 did not have the recovery task T9 in their data collection. Table \ref{tab:over} shows the data availability for the participants in each folder.  Each folder follows these naming conventions: The first letter of the filename shows the subject ID, followed by two letters showing the task number (e.g., T1). For example, LX corresponds to the validated stress levels of the task TX.

To assess stress levels during the stressful tasks, we administered self-assessment stress level questionnaires at 2-minute intervals. However, due to concerns about survey bias arising from the predictability of the questions \cite{hellevik2016extreme} when frequently completing both the NASA-TLX and self-assessment forms, we decided to have participants complete the NASA-TLX forms only once, at the end of the task, to measure the overall workload of the task. The labels (overall and intervals) are provided in the $stress\_level.csv$ files. Task T3 (delivering a presentation) lasted 5-9 minutes because some participants finished the task early. In these cases, we split the task duration into 2-minute segments for labeling, resulting in fewer labeled intervals for those participants who finished early.

\begin{table}[ht]
\caption{The availability of the tasks/data for each of the participants (\ding{55} mark corresponds to the case where the data is not available for the participant, \checkmark shows the availability of the data).}
\begin{tabular}{|l|l|l|l|l|l|l|l|l|l|l|l|l|l|l|l|l|}
\hline
\textbf{ID/Task} & \textbf{T1}               & \textbf{T2}               & \textbf{L2}                & \textbf{T3}               & \textbf{L3}                         & \textbf{T4}                        & \textbf{T5}                        & \textbf{L5}                         & \textbf{T6}                        & \textbf{T7}                        & \textbf{L7}                         & \textbf{T8}                        & \textbf{T9}                         & \textbf{Video}                      & \textbf{labels}                     & \textbf{Site}          \\ \hline
S1                & \checkmark & \checkmark & \ding{55} & \checkmark & \ding{55} & \checkmark & \checkmark & \ding{55} & \checkmark & \checkmark & \ding{55} & \checkmark & \checkmark  & \ding{55} & \ding{55} & Finland       \\ \hline
S2                & \checkmark & \checkmark & \ding{55} & \checkmark & \ding{55} & \checkmark & \checkmark & \ding{55} & \checkmark & \checkmark & \ding{55} & \checkmark & \checkmark  & \ding{55} & \ding{55} & Finland       \\ \hline
S3                & \checkmark & \checkmark & \ding{55} & \checkmark & \ding{55} & \checkmark & \checkmark & \ding{55} & \checkmark & \checkmark & \ding{55} & \checkmark & \checkmark  & \ding{55} & \ding{55} & Finland       \\ \hline
S4                & \checkmark & \checkmark & \ding{55} & \checkmark & \ding{55} & \checkmark & \checkmark & \ding{55} & \checkmark & \checkmark & \ding{55} & \checkmark & \checkmark  & \ding{55} & \ding{55} & Finland       \\ \hline
S5                & \checkmark & \checkmark & \ding{55} & \checkmark & \ding{55} & \checkmark & \checkmark & \ding{55} & \checkmark & \checkmark & \ding{55} & \checkmark & \checkmark  & \ding{55} & \ding{55} & Finland       \\ \hline
S6               & \checkmark & \checkmark & \ding{55} & \checkmark & \ding{55} & \checkmark & \checkmark & \ding{55} & \checkmark & \checkmark & \ding{55} & \checkmark & \checkmark  & \ding{55} & \ding{55} & Finland       \\ \hline
S7               & \checkmark & \checkmark & \checkmark  & \checkmark & \checkmark  & \checkmark & \checkmark & \checkmark  & \checkmark & \checkmark & \checkmark  & \checkmark & \ding{55} & \checkmark  & \checkmark  & U.S. \\ \hline
S8               & \checkmark & \checkmark & \checkmark  & \checkmark & \checkmark  & \checkmark & \checkmark & \checkmark  & \checkmark & \checkmark & \checkmark  & \checkmark & \ding{55} & \checkmark  & \checkmark  & U.S. \\ \hline
S9               & \checkmark & \checkmark & \checkmark  & \checkmark & \checkmark  & \checkmark & \checkmark & \checkmark  & \checkmark & \checkmark & \checkmark  & \checkmark & \ding{55} & \checkmark  & \checkmark  & U.S. \\ \hline
S10              & \checkmark & \checkmark & \checkmark  & \checkmark & \checkmark  & \checkmark & \checkmark & \checkmark  & \checkmark & \checkmark & \checkmark  & \checkmark & \ding{55} & \checkmark  & \checkmark  & U.S. \\ \hline
S11               & \checkmark & \checkmark & \checkmark  & \checkmark & \checkmark  & \checkmark & \checkmark & \checkmark  & \checkmark & \checkmark & \checkmark  & \checkmark & \ding{55} & \ding{55} & \checkmark  & U.S. \\ \hline
S12               & \checkmark & \checkmark & \checkmark  & \checkmark & \checkmark  & \checkmark & \checkmark & \checkmark  & \checkmark & \checkmark & \checkmark  & \checkmark & \ding{55} & \ding{55}  & \checkmark  & U.S. \\ \hline
S13               & \checkmark & \checkmark & \checkmark  & \checkmark & \checkmark  & \checkmark & \checkmark & \checkmark  & \checkmark & \checkmark & \checkmark  & \checkmark & \ding{55} & \ding{55} & \checkmark  & U.S. \\ \hline
S14              & \checkmark & \checkmark & \checkmark  & \checkmark & \checkmark  & \checkmark & \checkmark & \checkmark  & \checkmark & \checkmark & \checkmark  & \checkmark & \ding{55} & \ding{55} & \checkmark  & U.S. \\ \hline
S15              & \checkmark & \checkmark & \checkmark  & \checkmark & \checkmark  & \checkmark & \checkmark & \checkmark  & \checkmark & \checkmark & \checkmark  & \checkmark & \ding{55} & \ding{55} & \checkmark  & U.S. \\ \hline
S16              & \checkmark & \checkmark & \checkmark  & \checkmark & \checkmark  & \checkmark & \checkmark & \checkmark  & \checkmark & \checkmark & \checkmark  & \checkmark & \ding{55} & \ding{55} & \checkmark  & U.S. \\ \hline
S17              & \checkmark & \checkmark & \checkmark  & \checkmark & \checkmark  & \checkmark & \checkmark & \checkmark  & \checkmark & \checkmark & \checkmark  & \checkmark & \ding{55} & \checkmark& \checkmark  & U.S. \\ \hline
S18              & \checkmark & \checkmark & \checkmark  & \checkmark & \checkmark  & \checkmark & \checkmark & \checkmark  & \checkmark & \checkmark & \checkmark  & \checkmark & \ding{55} & \ding{55} & \checkmark  & U.S. \\ \hline
S19              & \checkmark & \checkmark & \checkmark  & \checkmark & \checkmark  & \checkmark & \checkmark & \checkmark  & \checkmark & \checkmark & \checkmark  & \checkmark & \ding{55} & \checkmark  & \checkmark  & U.S. \\ \hline
S20              & \checkmark & \checkmark & \checkmark  & \checkmark & \checkmark  & \ding{55} & \ding{55} & \ding{55}  & \ding{55} & \ding{55} & \ding{55}  & \ding{55} & \ding{55} & \ding{55} & \checkmark  & U.S. \\ \hline
S21               & \checkmark & \checkmark & \checkmark  & \checkmark & \checkmark  & \checkmark & \checkmark & \checkmark  & \checkmark & \checkmark & \checkmark  & \checkmark & \ding{55} & \checkmark  & \checkmark  & U.S.\\ \hline
S22                & \checkmark & \checkmark & \checkmark  & \checkmark & \checkmark  & \checkmark & \checkmark & \checkmark  & \checkmark & \checkmark & \checkmark  & \checkmark & \ding{55} & \checkmark  & \checkmark  & U.S. \\ \hline
S23              & \checkmark & \checkmark & \checkmark  & \checkmark & \checkmark  & \checkmark & \checkmark & \checkmark  & \checkmark & \checkmark & \checkmark  & \checkmark & \ding{55} & \checkmark  & \checkmark  & U.S. \\ \hline
S24               & \checkmark & \checkmark & \checkmark  & \checkmark & \checkmark  & \checkmark & \checkmark & \checkmark  & \checkmark & \checkmark & \checkmark  & \checkmark & \ding{55} & \ding{55}   & \checkmark  & U.S. \\ \hline
S25              & \checkmark & \checkmark & \checkmark  & \checkmark & \checkmark  & \checkmark & \checkmark & \checkmark  & \checkmark & \checkmark & \checkmark  & \checkmark & \ding{55} & \checkmark  & \checkmark  & U.S. \\ \hline
S26               & \checkmark & \checkmark & \checkmark  & \checkmark & \checkmark  & \checkmark & \checkmark & \checkmark  & \checkmark & \checkmark & \checkmark  & \checkmark & \ding{55} & \ding{55}   & \checkmark  & U.S. \\ \hline
S27              & \checkmark & \checkmark & \checkmark  & \checkmark & \checkmark  & \checkmark & \checkmark & \checkmark  & \checkmark & \checkmark & \checkmark  & \checkmark & \ding{55} & \ding{55} & \checkmark  & U.S. \\ \hline
S28               & \checkmark & \checkmark & \checkmark  & \checkmark & \checkmark  & \checkmark & \checkmark & \checkmark  & \checkmark & \checkmark & \checkmark  & \checkmark & \ding{55} & \checkmark  & \checkmark  & U.S. \\ \hline
S29               & \checkmark & \checkmark & \checkmark  & \checkmark & \checkmark  & \checkmark & \checkmark & \checkmark  & \checkmark & \checkmark & \checkmark  & \checkmark & \ding{55} & \checkmark  & \checkmark  & U.S. \\ \hline
S30               & \checkmark & \checkmark & \checkmark  & \checkmark & \checkmark  & \checkmark & \ding{55} & \ding{55}  & \ding{55} & \ding{55} & \ding{55}  & \ding{55} & \ding{55} & \ding{55}  & \checkmark  & U.S. \\ \hline
\end{tabular}
\label{tab:over}
\end{table}

\subsection*{Data files description}
We provide the files in the task sub-folders:
\begin{itemize}
    \item \textbf{Xception.xlsx}: The Excel file contains the facial expressions deduced at frame level for each task using MiniXception \cite{arriaga2017real}, trained over the FER-2013 dataset \cite{goodfellow2013challenges}. The details are given in Section "Facial expression recognition". The sampling rate is 30 frames per second.

    \item \textbf{HR.csv}: The file contains a single column with the average heart rate. The first row is the start time of the session in UTC time-stamp. The second row is the sampling rate expressed in Hz.

    \item \textbf{EDA.csv}: The file contains a single column with the average electrodermal activity. The first row is the start time of the session in UTC time-stamp. The second row is the sampling rate expressed in Hz.

    \item \textbf{TEMP.csv}: The file contains a single column with the average skin temperature. The first row is the start time of the session in UTC time-stamp. The second row is the sampling rate expressed in Hz.

    \item \textbf{Acc.csv} The file contains three columns corresponding to the x-axis, y-axis, and z-axis accelerometer data, respectively. The first row has the start times of the session in UTC time-stamp. The second row gives the sampling rates expressed in Hz.

    \item \textbf{V.mp4}: The video file contains raw video files of participants who consented to release their videos. This file is not available for all participants.

    \item \textbf{Stress\_level.csv}: Thefile holds the validated stress level of the participants extracted from their answers. Each label is for a two-minute interval (represented as 0, 1, 2, and unknown, where 0 = no-stress; 1=low-stress; 2=high-stress, unknown for unknown stress level).

    \item \textbf{Landmarks.csv}: The file contains the normalized position of the participant's 68 facial landmarks (Dlib features described in Section "Dlib features") over time. The sampling rate of the facial landmarks is 30 frames per second, which is not provided in the landmarks csv file.
    
    \item \textbf{BVP.csv}: The file contains a single column with the blood volume pulse signal. The first row is the start time of the session in UTC time-stamp. The second row is the sampling rate expressed in Hz.

    \item \textbf{IBI.csv}: The file contains two columns; the first column shows the time-stamp, and the second column contains the corresponding inter-beat interval values. 

\end{itemize}

The videos and physiological signals have exact start times and can be synced by interpolation, downsampling, or upsampling data processing methods.

\section*{Technical validation}
For the analysis presented in this paper, only subjects with complete and labeled datasets were considered. As a result, subjects S1 to S6, S20, and S30 were excluded. This exclusion was necessary to ensure the reliability and validity of the results, as missing labels or incomplete datasets could introduce bias or inaccuracies in the analysis.  We used multi-class macro-averaging that calculates the metric independently for each stress level (low-, medium-, and high-stress) and then takes the average, treating all classes equally.

% \subsection*{Metrics}
% In this paper, we compare the performance of classification methods using the following metrics:

% \begin{itemize}

% \item \textbf{Precision} represents the closeness of the predicted results:
%     \begin{equation}
%         Precision = \frac{TP}{TP + FP}.
%     \end{equation}
%     where the TP is the number of True Positives and FP is the number of False Positives.
    
% \item\textbf{Recall} represents the fraction of correctly predicted positive observations to all actual class samples:
%     \begin{equation}
%         recall= \frac{TP}{TP + FN}.
%     \end{equation}

% \item\textbf{f1\_score} combines the precision and recall of a classifier into a single metric by taking their harmonic mean:
%     \begin{equation}
%         f1\_score = \frac{2\times Precision \times Recall}{Precision + Recall}.
%     \end{equation}
% \end{itemize}

\subsection*{Facial Action Units}

In this paper, we investigated the correlations between stress level and frequency of the different AUs in a time window of 20 seconds among 24 subjects with self-reported stress level labels. Table \ref{tab:au} shows the incidence of stress manifestation through facial AUs across the subjects. We generated AUs from facial landmarks \cite{hjortsjo1969man} and compared the scores during different stress levels. It is noteworthy that not all subjects exhibited identical AUs when exposed to stress \cite{nicolle2012robust}. Despite this variability, we identified and reported correlations between specific AUs and stress in certain individuals, with detailed results presented in Table \ref{tab:au}.

\begin{table}[htbp]
\centering
\caption{Distribution of Facial Action Units indicative of stress among 24 subjects with self-reported stress level labels}
\begin{tabular}{|l|l|l|} \hline
Action Unit         &AU number                       & Number of subjects \\ \hline
Lip Tightener       &12                     & 17                 \\ \hline
Lip Stretcher       &11                     & 16                 \\ \hline
Mouth Stretch       &15                     & 15                 \\ \hline
Lip Depressor       &8                      & 14                 \\ \hline
Lid Droop           &25                     & 13                 \\ \hline
Nostril Compressor  &24                     & 10                 \\ \hline
Cheek Raiser        &4                      & 9                  \\ \hline
\end{tabular}
\label{tab:au}
\end{table}

\subsection*{Machine learning and stress detection model}
The EDA, HR, Temp, ACC signals, and video were collected at different sampling rates due to the variation of sensors. The frequency of the signals ranges from 1 to 30 Hz for heart rate and video data, respectively. We used the physiological signals, facial expressions, and AUs generated from videos as features for our stress detection model. We decided to use a frequency of 4 Hz after evaluating the accuracy of different models to minimize the information loss while monitoring the computational cost of the models. The code associated with the data preprocessing, upsampling, and downsampling is available on the GitHub repository  \cite{Majid2022}. The machine learning classification models process the physiological signals and facial expression features. The stress detection models are trained and tested based on the participants' survey answers. The output of the stress detection models is the stress level of the participant at the mentioned time step (0: No stress, 1: medium stress, 2: high stress).

\subsection*{Physiological signals vs. facial expressions}

\begin{figure}[htbp]
    \centering
    \includegraphics[width=.6\textwidth]{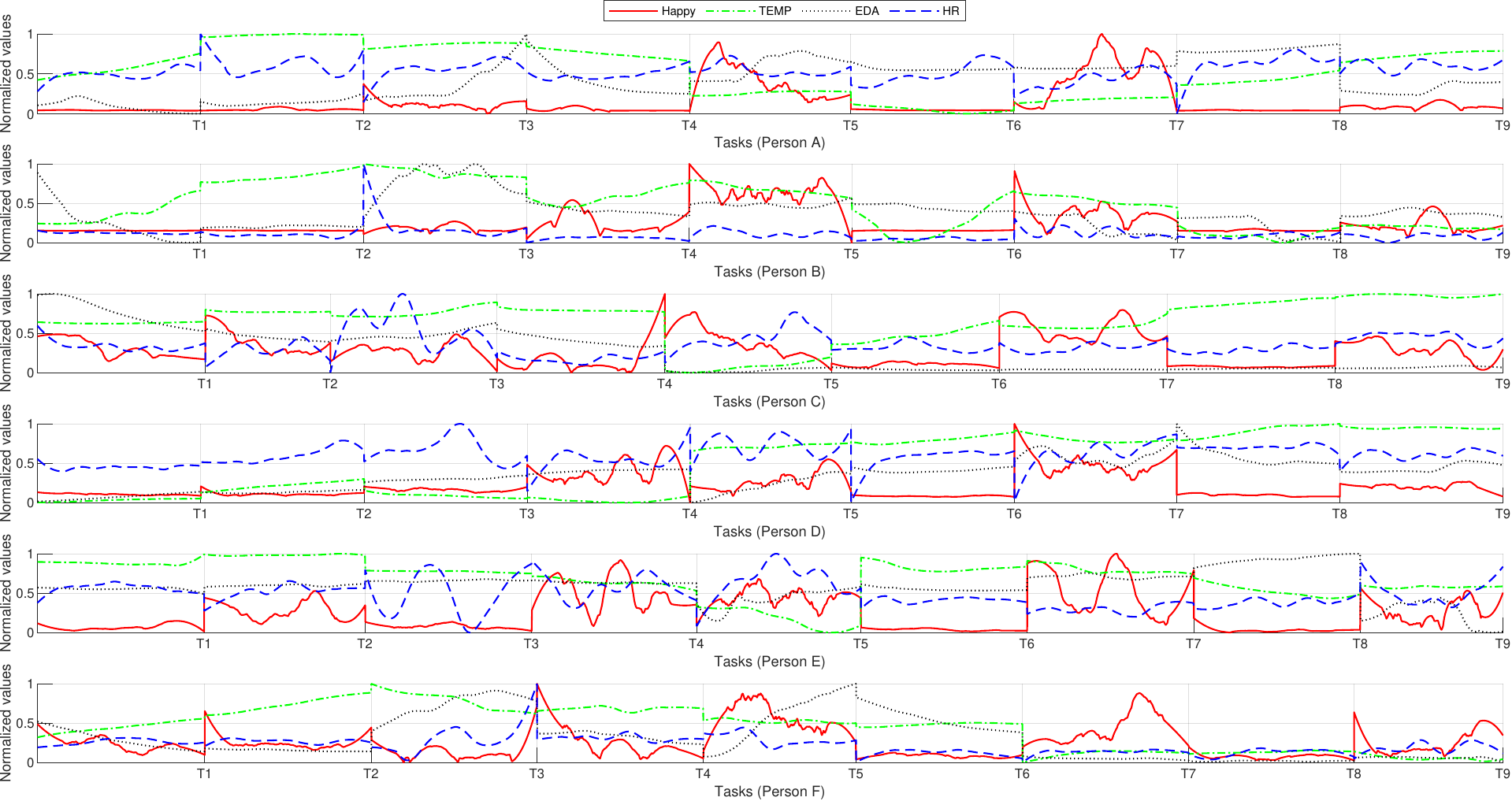}
    \caption{Analysis of subjects S1-S6 physiological signals and facial expressions for tasks T1-T9.}
\label{subjectA}
\end{figure}

We analyzed the physiological signals obtained from the Empatica E4 watch and the cla facial expression probabilities from the video data for different tasks. Since the frame rate for the video is higher than the frequency of the physiological signals, we interpolated the physiological data to make it equal in length to the length of the deduced emotions from the video frames. To smooth the emotion data (Figure \ref{subjectA}), we use a quadratic polynomial used in the Savitzky-Golay method \cite{schafer2011savitzky}. The smoothing filter was applied to the data at the task level by keeping the span of the smoothing filter at 30 percent. After applying the smoothing filter, we scaled the data between 0 and 1 by applying $\tilde{x_i} = \frac{x_i - x_{min}}{x_{max}-x_{min}}$, where $\tilde{x_i}$ is the scaled value. $x_{min}$ is the minimum, and $x_{max}$ is the maximum smoothed value at the corresponding task/session for the participant.

Figure \ref{subjectA} shows the happiness curves and physiological signals of subjects S1-S6 performing different tasks (T1-T9). The happiness curve is observed to fluctuate more in T5 and T7 for subjects S1-S6 as compared to other tasks. For T5, the fluctuation is attributed to the cognitive load and conflicting stimuli inherent in the task at hand, which evoke momentary frustration or challenge followed by happiness upon correct responses, resulting in emotional variability \cite{etkin2011emotional}. The fluctuations observed during T7 likely reflect the dynamic, amusing content of the videos themselves, which may induce peaks of amusement incorporated with neutral or less engaging moments \cite{wang2015video}. For a given task, the temperature is observed to stay steady within the task; however, some differences have been observed in the temperature values between the tasks. The EDA curve is observed to stay steady during T1 and T2, while a slight variation is observed in T3. We also observed a peak heart rate value for subjects S2, S3, and S4 for T3. Although these curves illustrate how each task uniquely modulates facial expression and physiological signals, they do not establish causal relationships or explain why those modulations occur. Detailed modeling or controlled follow-up studies will be required to find the patterns of these observed fluctuations.
Similar plots can be generated for comparing other emotions (Angry, Disgust, Fear, Sad, Surprise) to the physiological signals provided in the dataset.

\begin{figure}[htbp]
    \centering
    \includegraphics[width=.5\linewidth]{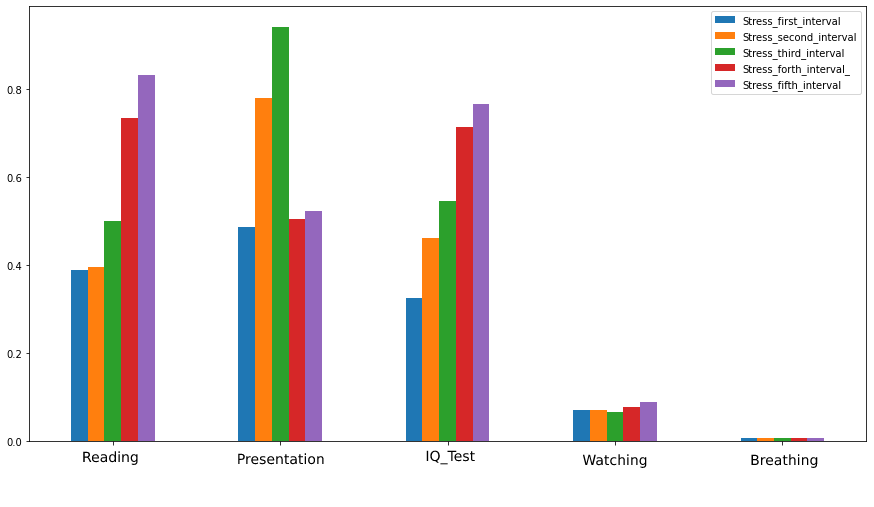}
    \caption{Overall stress level of the students performing different stressful tasks}
    \label{fig:stacked}
\end{figure}

Figure \ref{fig:stacked} shows the overall stress level of the students S7-S30 performing different tasks based on participants' feedback. For example, the overall stress level in delivering a presentation was higher than in the other tasks. The stress level of the task in the presentation had a peak exactly in the middle of the presentation, compared to reading and IQ tests, where the overall stress level of the subjects increased while reaching the end of the task. In addition, the stress level during preparation for a presentation showed similar results to those of taking a test for the participants.

\subsubsection*{Feature Extraction:}
A stress detection algorithm was designed to detect stress from facial expressions and physiological signals. In order to extract the features for the stress detection model, we first performed facial expression recognition in the video as described in the Section "Facial expression recognition". Consistent with prior work on wearable stress classification
\cite{gjoreski2017continuous,hernandez2011using,kim2018stress}, we extracted features in 20-second windows with a 5-second step (75\% overlap). Twenty seconds is long enough to capture several phasic EDA responses \cite{benedek2010continuous} and to stabilize HRV \cite{mccraty2015heart} statistics while remaining short enough to detect rapid stress onsets \cite{Boucsein2012Electrodermal,shaffer2017overview}. In total, we used 28 features extracted from both video and physiological signals.
\begin{itemize}
    \item \textbf{Video signal features:}
    
    The mean of the facial expressions extracted from the video was concatenated with the features extracted from biometric signals. For each of the 3 physiological signals, we calculated the mean, minimum, maximum, and standard deviation along the 20-second sliding window. 
    \item \textbf{Physiological signals features:}

    In addition, we extracted the kurtosis, skewness, number of peaks, amplitude, and duration of the peak for the EDA signal. For HR, we extracted the root mean square (quadratic mean), number of peaks, amplitude, and duration of the peaks. These statistical features for physiological signals were extracted based on an earlier work by \cite{el2019random}. 

    \item \textbf{Labels:}

    The labels for each window were calculated based on the average stress values during the time window: "no stress" when $S \le 0.65$, "medium stress" when $0.65 < S \le1.3$, and "high stress" when $S > 1.3$. The source code for the stress detection algorithm comprises of feature extraction, stress detection, and change-point detection, and is provided in the GitHub repository \cite{Majid2022}. 
\end{itemize}

Three models (Random Forest, Decision Tree, and XGBoost) were trained from the features extracted in the second step to predict stress during the 20-second window. To evaluate the effectiveness of the model and avoid overfitting, we followed a leave-one-subject-out strategy on the test dataset, where the model is trained on data from several participants and evaluated against the remainder. This process is repeated for all the participants. This dataset is imbalanced, and stress levels are distributed as: 1- low stress: $76\%$, 2- medium stress: $16\%$, and 3- high stress: $8\%$. We employed the SMOTE oversampling approach to balance the data \cite{chawla2002smote}. The precision, recall, and F-score of these three approaches are presented in Table \ref{tab:clfloso}.

\begin{table}[htbp]
\centering
\begin{tabular}{|lclll|}
\hline
\multicolumn{1}{|l|}{\textbf{Model}}               & \multicolumn{1}{l|}{\textbf{Features}} & \multicolumn{1}{l|}{\textbf{Recall}}                                                                  & \multicolumn{1}{l|}{\textbf{Precision}}           & \textbf{F1\_score}                                                               \\ \hline
\multicolumn{5}{|c|}{\textbf{Random Forest}}                                                                                                                                                                                                                                                                                               \\ \hline
\multicolumn{1}{|l|}{\textbf{Emotions}}            & \multicolumn{1}{c|}{7}                 & \multicolumn{1}{l|}{48.64}                                                                            & \multicolumn{1}{l|}{50.08} & 48.02                                                                            \\ \hline
\multicolumn{1}{|l|}{\textbf{Biometric}}           & \multicolumn{1}{c|}{21}                & \multicolumn{1}{l|}{65.62}                                                                            & \multicolumn{1}{l|}{64.07}                        & 61.18                                                                            \\ \hline
\multicolumn{1}{|l|}{\textbf{Multimodal (Top 10)}} & \multicolumn{1}{c|}{10}                & \multicolumn{1}{l|}{\textbf{79.72}}                                                                   & \multicolumn{1}{l|}{\textbf{73.46}}               & \textbf{69.32}                                                                   \\ \hline
\multicolumn{5}{|c|}{\textbf{Decision Tree}}                                                                                                                                                                                                                                                                                               \\ \hline
\multicolumn{1}{|l|}{\textbf{Emotions}}            & \multicolumn{1}{c|}{7}                 & \multicolumn{1}{l|}{45.56}          & \multicolumn{1}{l|}{46.13}                        & 45.51                                                                            \\ \hline
\multicolumn{1}{|l|}{\textbf{Biometric}}           & \multicolumn{1}{c|}{21}                & \multicolumn{1}{l|}{ 58.25} & \multicolumn{1}{l|}{58.49}               & {56.37} \\ \hline
\multicolumn{1}{|l|}{\textbf{Top 10}}              & \multicolumn{1}{c|}{10}                & \multicolumn{1}{l|}{\textbf{69.89}}                                                                            & \multicolumn{1}{l|}{\textbf{63.86}}                        & \textbf{60.19}                                                                            \\ \hline
\multicolumn{5}{|c|}{\textbf{XGBoost}}                                                                                                                                                                                                                                                                                                     \\ \hline
\multicolumn{1}{|l|}{\textbf{Emotions}}            & \multicolumn{1}{c|}{7}                 & \multicolumn{1}{l|}{70.52}                                                     & \multicolumn{1}{l|}{ 69.69} & { 66.44}                                                     \\ \hline
\multicolumn{1}{|l|}{\textbf{Biometric}}           & \multicolumn{1}{c|}{21}                & \multicolumn{1}{l|}{\textbf{79.30}}                                                     & \multicolumn{1}{l|}{ 74.28} & {\textbf{ 70.16}}                                                     \\ \hline
\multicolumn{1}{|l|}{\textbf{Multimodal (Top 10)}} & \multicolumn{1}{c|}{10}                & \multicolumn{1}{l|}{\textbf{81.64}}                                                                   & \multicolumn{1}{l|}{\textbf{74.82}}               & 69.28                                                                   \\ \hline
\end{tabular}
\caption{Performance of different models using balanced data and Leave One Subject Out}
\label{tab:clfloso}
\end{table}

 The emotions model used the mean of the facial expressions extracted from the video to detect stress. The biometrics model used the 21 biometric features to train the machine learning model. A large number of features may lead to an overfitted model that is not generalizable to unseen data (curse of dimensionality) \cite{salam2021effect}. To reduce the number of features, we used Pearson correlation analysis \cite{benesty2009pearson} to select the top 10 features correlated with stress. The top 10 features we observed were 'lip\_puckerer\_mean', 'lower\_lip\_depressor\_mean', 'temp\_mean', 'temp\_max', 'temp\_min', 'eda\_mean', 'eda\_max', 'eda\_min', 'eda\_peak\_amplitude', and 'hr\_min' using information gain. We trained each machine learning model with these 10 features to predict the value of stress. The code for training the machine learning models and evaluating the results is available in GitHub \cite{Majid2022}.

We evaluated the performance of Random Forest, Decision Tree, and XGBoost, using balanced data and a Leave-One-Subject-Out (LOSO) cross-validation. The models were trained using facial expression, physiological signals, and a multimodal combination of the top ten features. The Random Forest classifier, with 100 trees (n\_estimators=100), a depth of 7(max\_depth=7), and a minimum sample leaf size of 5, achieved the best performance with the multimodal feature set, with a recall of 79.72\%, precision of 73.46\%, and an F1-score of 69.32\% compared to emotions alone with a recall of 48.64\% and an F1-score of 48.02\%. The Decision Tree classifier, using a maximum depth of 5 and a minimum sample leaf size of 3, with the multimodal features, achieved a recall of 69.89\%, precision of 63.86\%, and an F1-score of 60.19\%, while its performance with emotion signals alone remained limited, with a recall of 45.56\% and an F1-score of 45.51\%. The XGBoost model, with a maximum depth of 6 and a gamma value of 0, outperformed other models with the biometric feature set, achieving a recall of 79.30, a precision of 74.28, and an F1-score of 70.16.

Emotions, such as anger (angry), surprise, and sadness (sad), exhibit high correlation with stress \cite{wang2020heightened,giannakakis2020automatic, McDuff2014RemoteStress}. Several authors have previously discussed this relationship between heart rate, emotions, and stress \cite{taelman2009influence, vrijkotte2000effects, schubert2009effects, goodie2000validation}. Earlier studies on stress detection evaluated the relationship between stress and skin temperature \cite{chen2021pain, li2020stress}. The skin temperature drops during the onset of stress \cite{zhai2006stress} and increases above the normal temperature after the onset \cite{yamakoshi2008feasibility,zhai2006stress}. Therefore, skin temperature and stress are not directly correlated. However, a sudden decrease in skin temperature is a good indicator of early stress, but during later time periods, the skin temperature is higher. The machine learning models can use these complex relationships to identify high-stress levels, as represented in Table \ref{tab:clfloso}.

\subsection*{Linear Models with Mixed Effects}
Linear Models with Mixed effects (LMMs) are an extension of linear models that incorporate fixed and random effects. These models are especially useful for hierarchical or nested data. 

\begin{itemize}
    \item \textbf{Mixed effects} in LMMs are analogous to the coefficients in standard linear regression models, representing the average effect of explanatory variables on the response variable. 
    \item \textbf{Random effects} account for variations at different levels of the hierarchy, such as differences between subjects or groups. This dual structure allows LMMs to model the overall trend and the individual deviations from this trend.
\end{itemize}

\begin{table}[htbp]
\centering
\caption{Mixed linear models intercept results of different signals in different stress groups using lbfgs method}\label{tab:FFFlem}

\begin{tabular}{|l|l|l|l|l|l|l|}
\hline
\textbf{Feature}                            & \textbf{Coef.}  & \textbf{Std.Err.} & \textbf{Z}      & \textbf{P\textgreater{}|Z|} & \textbf{{[}0.025} & \textbf{0.975{]}} \\ \hline
\textbf{hr\_std                           } & -0.111 & 0.035    & -3.163 & 0.002              & -0.180   & -0.042   \\ \hline
\textbf{hr\_peak\_amplitude               } & 0.038  & 0.017    & 2.299  & 0.021              & 0.006    & 0.071    \\ \hline
\textbf{hr\_peak\_duration                } & -0.020 & 0.007    & -2.855 & 0.004              & -0.034   & -0.006   \\ \hline
\textbf{eda\_peak\_duration               } & 0.007  & 0.003    & 2.249  & 0.025              & 0.001    & 0.013    \\ \hline
\textbf{inner\_brow\_raiser\_mean         } & -0.058 & 0.018    & -3.216 & 0.001              & -0.093   & -0.023   \\ \hline
\textbf{inner\_brow\_raiser\_max          } & -0.046 & 0.017    & -2.725 & 0.006              & -0.079   & -0.013   \\ \hline
\textbf{inner\_brow\_raiser\_kurtosis     } & 0.007  & 0.003    & 2.053  & 0.040              & 0.000    & 0.013    \\ \hline
\textbf{inner\_brow\_raiser\_skewness     } & 0.007  & 0.004    & 1.968  & 0.049              & 0.000    & 0.015    \\ \hline
\textbf{brow\_lowerer\_mean               } & -0.060 & 0.029    & -2.095 & 0.036              & -0.117   & -0.004   \\ \hline
\textbf{upper\_lid\_raiser\_min           } & 0.076  & 0.038    & 2.006  & 0.045              & 0.002    & 0.151    \\ \hline
\textbf{upper\_lid\_raiser\_max           } & -0.091 & 0.036    & -2.555 & 0.011              & -0.160   & -0.021   \\ \hline
\textbf{upper\_lid\_raiser\_std           } & 0.051  & 0.022    & 2.272  & 0.023              & 0.007    & 0.095    \\ \hline
\textbf{upper\_lid\_raiser\_skewness      } & 0.020  & 0.010    & 2.073  & 0.038              & 0.001    & 0.039    \\ \hline
\textbf{cheek\_raiser\_feature\_min       } & 0.061  & 0.026    & 2.362  & 0.018              & 0.010    & 0.111    \\ \hline
\textbf{lid\_tightener\_mean              } & 0.141  & 0.055    & 2.583  & 0.010              & 0.034    & 0.249    \\ \hline
\textbf{lid\_tightener\_std               } & -0.066 & 0.021    & -3.058 & 0.002              & -0.108   & -0.024   \\ \hline
\textbf{nose\_wrinkler\_std               } & -0.033 & 0.012    & -2.719 & 0.007              & -0.057   & -0.009   \\ \hline
\textbf{nose\_wrinkler\_kurtosis          } & 0.008  & 0.004    & 1.982  & 0.047              & 0.000    & 0.017    \\ \hline
\textbf{upper\_lip\_raiser\_min           } & -0.031 & 0.013    & -2.376 & 0.018              & -0.056   & -0.005   \\ \hline
\textbf{upper\_lip\_raiser\_kurtosis      } & -0.012 & 0.004    & -3.291 & 0.001              & -0.019   & -0.005   \\ \hline
\textbf{nasolabial\_furrow\_deepener\_mean} & 0.049  & 0.025    & 1.983  & 0.047              & 0.001    & 0.097    \\ \hline
\textbf{dimpler\_std                      } & 0.029  & 0.014    & 2.040  & 0.041              & 0.001    & 0.058    \\ \hline
\textbf{lip\_corner\_depressor\_mean      } & 0.085  & 0.025    & 3.339  & 0.001              & 0.035    & 0.135    \\ \hline
\textbf{lip\_corner\_depressor\_max       } & 0.056  & 0.022    & 2.525  & 0.012              & 0.013    & 0.100    \\ \hline
\textbf{lip\_corner\_depressor\_kurtosis  } & -0.011 & 0.003    & -3.336 & 0.001              & -0.018   & -0.005   \\ \hline
\textbf{lower\_lip\_depressor\_min        } & 0.241  & 0.076    & 3.190  & 0.001              & 0.093    & 0.389    \\ \hline
\textbf{lower\_lip\_depressor\_skewness   } & -0.007 & 0.003    & -2.011 & 0.044              & -0.014   & -0.000   \\ \hline
\textbf{chin\_raiser\_max                 } & -0.088 & 0.027    & -3.255 & 0.001              & -0.140   & -0.035   \\ \hline
\textbf{chin\_raiser\_std                 } & -0.048 & 0.014    & -3.280 & 0.001              & -0.076   & -0.019   \\ \hline
\textbf{lip\_puckerer\_kurtosis           } & 0.009  & 0.004    & 2.093  & 0.036              & 0.001    & 0.017    \\ \hline
\textbf{lip\_funneler\_max                } & 0.054  & 0.023    & 2.384  & 0.017              & 0.010    & 0.099    \\ \hline
\textbf{lip\_funneler\_skewness           } & 0.013  & 0.005    & 2.612  & 0.009              & 0.003    & 0.023    \\ \hline
\textbf{jaw\_drop\_min                    } & -0.066 & 0.030    & -2.229 & 0.026              & -0.124   & -0.008   \\ \hline
\textbf{jaw\_thrust\_skewness             } & -0.040 & 0.017    & -2.387 & 0.017              & -0.073   & -0.007   \\ \hline
\textbf{puff\_std                         } & -0.030 & 0.013    & -2.382 & 0.017              & -0.055   & -0.005   \\ \hline
\textbf{tongue\_bulge\_std                } & -0.032 & 0.013    & -2.548 & 0.011              & -0.057   & -0.007   \\ \hline
\textbf{nostil\_dilator\_std              } & 0.027  & 0.012    & 2.199  & 0.028              & 0.003    & 0.052    \\ \hline
\textbf{lid\_droop\_mean                  } & 0.037  & 0.013    & 2.821  & 0.005              & 0.011    & 0.063   \\ \hline
\end{tabular}
\end{table}

To handle potential outliers, we applied user-level standardization. Table \ref{tab:FFFlem} presents the LMM results for physiological features and facial AUs when participants were treated as random effects. For example, the feature hr\_std has a significant negative correlation (p<0.05) to high stress, b = -0.111, z = -3.163, p = 0.002. Similarly, inner\_brow\_raiser\_mean has a negative coefficient of -0.058 (p = 0.001), illustrating that higher inner brow-raiser activity is associated with low-stress levels. We also observe positive relationships for lid\_tightener\_mean (coefficient = 0.141, p = 0.010) and lip\_corner\_depressor\_mean (coefficient = 0.085, p = 0.001) with higher stress levels. Multimodal features in the form of facial AUs and physiological signals together showed statistically significant correlations that are predictive for stress modeling.

For model fitting, we used the Limited-memory Broyden-Fletcher-Goldfarb-Shanno optimization algorithm, method='lbfgs', which is quite efficient for large datasets with complex random effect structures. We also have robust maximum likelihood estimates. These findings point to LMM's utility for modeling with hierarchical data on stress, while accounting for the variation across both individual and group levels.

\section* {Limitations and future works}

We collected data in laboratory conditions, and the following items are the limitations of our stress detection dataset:
\begin{itemize}

    \item Not all of the dataset is covered by self-reported stress level labels. Unlabeled data do not necessarily imply a lack of stress; it just means that we did not decide to collect the labels in assumed low-stress tasks. However, we asked participants to let us know if they experienced the onset of stress in low-stress tasks. Except for one participant, none of the subjects experienced stressful moments in the low-stress tasks.
    \item The self-reported stress levels have a granularity of 2 minutes. Finer granularity can lead to recall bias \cite{tarrant1993effects}.
    \item This dataset is primarily focused on labeling acute stress among participants and does not consider chronic stress as a factor in the participants. There is no sufficient research on chronic stress detection using physiological signals. 
    \item We conducted a laboratory setting for stress detection, and the participants may experience some stress due to the new environment. However, we asked the participants to read magazines under task 1 to mitigate the newcomer's socialization through the Lens of Stress effect \cite{allen1999newcomer}.
    \item Our study primarily measures arousal rather than stress, as the tasks involved do not incorporate high stakes or a significant audience. Future research could benefit from incorporating these elements to elicit genuine responses to stress.
\end{itemize}
% \section*{Usage notes}
% % Let's have a discussion on how we arrived at this uses
% The EmpathicSchool dataset can be used for various applications. We envision the following potential applications and further analyses for the following research communities.\\\\
% \textbf{Stress management:}
% The EmpathicSchool dataset can be used to analyze human facial expressions in correlation with stress. The stress management applications help improve humans' everyday functioning and mitigate chronic stress.\\\\
% \textbf{Emphatic building:}
% The EmpathicSchool dataset consists of data at different stress levels (sessions), which are analogous to different events in daily life in the working space. The dataset is a stepping stone toward analyzing employees' well-being in an Emphatic building environment.\\\\
% \textbf{Perceptual interface:}
% The dataset is also useful for further studies in developing perceptual interfaces for patients in the Intensive Care Unit (ICU). Analyzing human expressions in co-relation with physiological signals can be useful for pain management, and similar protocols can be used for enhancing communication, reducing stress, and improving cognitive function in critically stressful environments.\\\\
% \textbf{Online education:}
% In online education, it is important to adapt the style and comfort according to the student's needs. To render students satisfaction, frustration, or stress levels, the EmpathicSchool dataset can be used to develop such applications.

\section*{Code availability}
We provide the code for deducing facial expressions from video data using the InceptionV3 \cite{szegedy2016rethinking} model pre-trained on the FER-2013 dataset \cite{goodfellow2013challenges}, and dlib features \cite{king2009dlib} that provided 68 facial landmarks on a facial image as discussed in Section "Dlib feature". We also provide codes for analyzing (producing the plots/curves) physiological signals vs. facial expressions. The code is publicly available on GitHub \cite{Majid2022} and the data is publicly available on Zenodo \cite{ZenodoDataset}.

\section*{Acknowledgements}
This work has been supported by NSF IUCRC CVDI, project AMALIA funded by Business Finland and DSB, as well as projects Mad@work and Stroke-Data funded by Haltian. 

\section*{Authors contributions}
Conceptualization: R.G., M.G., A.I, S.K., R.B. and J.R.; Methodology: M.H., F.S., S.K and R.B.; Software: M.H., F.S., S.K.,
and R.B.; Validation: M.H., F.S., S.K. and R.B., A.I., J.R., M.G., and R.G.; Formal analysis: M.H. F.S.;
Investigation: M.H., F.S.; Resources: R.G., and M.G.; Writing—original draft preparation: M.H., F.S.; Writing—review and editing: R.G., S.K, R.B, A.I., J.R. and
M.G.; Visualization: F.S. and M.H.; Supervision: R.G., A.I., S.K., R.B., J.R.,  and M.G.; Project administration:
R.G. and M.G.; Funding acquisition: R.G. and M.G. All authors reviewed the manuscript.
\section*{Competing Interests}
The authors declare no competing interests.
\bibliography{main}

\begin{thebibliography}{100}
\urlstyle{rm}
\expandafter\ifx\csname url\endcsname\relax
  \def\url#1{\texttt{#1}}\fi
\expandafter\ifx\csname urlprefix\endcsname\relax\def\urlprefix{URL }\fi
\expandafter\ifx\csname doiprefix\endcsname\relax\def\doiprefix{DOI: }\fi
\providecommand{\bibinfo}[2]{#2}
\providecommand{\eprint}[2][]{\url{#2}}

\bibitem{poria2017review}
\bibinfo{author}{Poria, S.}, \bibinfo{author}{Cambria, E.},
  \bibinfo{author}{Bajpai, R.} \& \bibinfo{author}{Hussain, A.}
\newblock \bibinfo{journal}{\bibinfo{title}{A review of affective computing:
  From unimodal analysis to multimodal fusion}}.
\newblock {\emph{\JournalTitle{Information Fusion}}}
  \textbf{\bibinfo{volume}{37}}, \bibinfo{pages}{98--125}
  (\bibinfo{year}{2017}).

\bibitem{can2019stress}
\bibinfo{author}{Can, Y.~S.}, \bibinfo{author}{Arnrich, B.} \&
  \bibinfo{author}{Ersoy, C.}
\newblock \bibinfo{journal}{\bibinfo{title}{Stress detection in daily life
  scenarios using smart phones and wearable sensors: A survey}}.
\newblock {\emph{\JournalTitle{Journal of biomedical informatics}}}
  \textbf{\bibinfo{volume}{92}}, \bibinfo{pages}{103139}
  (\bibinfo{year}{2019}).

\bibitem{plarre2011continuous}
\bibinfo{author}{Plarre, K.} \emph{et~al.}
\newblock \bibinfo{title}{Continuous inference of psychological stress from
  sensory measurements collected in the natural environment}.
\newblock In \emph{\bibinfo{booktitle}{Proceedings of the 10th ACM/IEEE
  international conference on information processing in sensor networks}},
  \bibinfo{pages}{97--108} (\bibinfo{organization}{IEEE},
  \bibinfo{year}{2011}).

\bibitem{smets2018large}
\bibinfo{author}{Smets, E.} \emph{et~al.}
\newblock \bibinfo{journal}{\bibinfo{title}{Large-scale wearable data reveal
  digital phenotypes for daily-life stress detection}}.
\newblock {\emph{\JournalTitle{NPJ digital medicine}}}
  \textbf{\bibinfo{volume}{1}}, \bibinfo{pages}{1--10} (\bibinfo{year}{2018}).

\bibitem{wijsman2011towards}
\bibinfo{author}{Wijsman, J.}, \bibinfo{author}{Grundlehner, B.},
  \bibinfo{author}{Liu, H.}, \bibinfo{author}{Hermens, H.} \&
  \bibinfo{author}{Penders, J.}
\newblock \bibinfo{title}{Towards mental stress detection using wearable
  physiological sensors}.
\newblock In \emph{\bibinfo{booktitle}{2011 Annual International Conference of
  the IEEE Engineering in Medicine and Biology Society}},
  \bibinfo{pages}{1798--1801} (\bibinfo{organization}{IEEE},
  \bibinfo{year}{2011}).

\bibitem{karikoski2013contextual}
\bibinfo{author}{Karikoski, J.} \& \bibinfo{author}{Soikkeli, T.}
\newblock \bibinfo{journal}{\bibinfo{title}{Contextual usage patterns in
  smartphone communication services}}.
\newblock {\emph{\JournalTitle{Personal and ubiquitous computing}}}
  \textbf{\bibinfo{volume}{17}}, \bibinfo{pages}{491--502}
  (\bibinfo{year}{2013}).

\bibitem{politou2017survey}
\bibinfo{author}{Politou, E.}, \bibinfo{author}{Alepis, E.} \&
  \bibinfo{author}{Patsakis, C.}
\newblock \bibinfo{journal}{\bibinfo{title}{A survey on mobile affective
  computing}}.
\newblock {\emph{\JournalTitle{Computer Science Review}}}
  \textbf{\bibinfo{volume}{25}}, \bibinfo{pages}{79--100}
  (\bibinfo{year}{2017}).

\bibitem{wilson2009fear}
\bibinfo{author}{Wilson, G.} \& \bibinfo{author}{Mossakowski, K.}
\newblock \bibinfo{journal}{\bibinfo{title}{Fear of job loss: Racial/ethnic
  differences in privileged occupations}}.
\newblock {\emph{\JournalTitle{Du Bois Review: Social Science Research on
  Race}}} \textbf{\bibinfo{volume}{6}}, \bibinfo{pages}{357--374}
  (\bibinfo{year}{2009}).

\bibitem{peternel2012presence}
\bibinfo{author}{Peternel, K.}, \bibinfo{author}{Poga{\v{c}}nik, M.},
  \bibinfo{author}{Tav{\v{c}}ar, R.} \& \bibinfo{author}{Kos, A.}
\newblock \bibinfo{journal}{\bibinfo{title}{A presence-based context-aware
  chronic stress recognition system}}.
\newblock {\emph{\JournalTitle{Sensors}}} \textbf{\bibinfo{volume}{12}},
  \bibinfo{pages}{15888--15906} (\bibinfo{year}{2012}).

\bibitem{bickford2005stress}
\bibinfo{author}{Bickford, M.}
\newblock \bibinfo{journal}{\bibinfo{title}{Stress in the workplace: A general
  overview of the causes, the effects, and the solutions}}.
\newblock {\emph{\JournalTitle{Canadian Mental Health Association Newfoundland
  and Labrador Division}}} \textbf{\bibinfo{volume}{8}}, \bibinfo{pages}{1--3}
  (\bibinfo{year}{2005}).

\bibitem{wellen2005inflammation}
\bibinfo{author}{Wellen, K.~E.}, \bibinfo{author}{Hotamisligil, G.~S.}
  \emph{et~al.}
\newblock \bibinfo{journal}{\bibinfo{title}{Inflammation, stress, and
  diabetes}}.
\newblock {\emph{\JournalTitle{The Journal of clinical investigation}}}
  \textbf{\bibinfo{volume}{115}}, \bibinfo{pages}{1111--1119}
  (\bibinfo{year}{2005}).

\bibitem{greenglass2001workload}
\bibinfo{author}{Greenglass, E.~R.}, \bibinfo{author}{Burke, R.~J.} \&
  \bibinfo{author}{Fiksenbaum, L.}
\newblock \bibinfo{journal}{\bibinfo{title}{Workload and burnout in nurses}}.
\newblock {\emph{\JournalTitle{Journal of community \& applied social
  psychology}}} \textbf{\bibinfo{volume}{11}}, \bibinfo{pages}{211--215}
  (\bibinfo{year}{2001}).

\bibitem{chu2024physiology}
\bibinfo{author}{Chu, B.}, \bibinfo{author}{Marwaha, K.},
  \bibinfo{author}{Sanvictores, T.}, \bibinfo{author}{Awosika, A.~O.} \&
  \bibinfo{author}{Ayers, D.}
\newblock \bibinfo{title}{Physiology, stress reaction}.
\newblock In \emph{\bibinfo{booktitle}{StatPearls [Internet]}}
  (\bibinfo{publisher}{StatPearls Publishing}, \bibinfo{year}{2024}).

\bibitem{filipovic2001review}
\bibinfo{author}{Filipovic, S.~R.}
\newblock \bibinfo{journal}{\bibinfo{title}{Review of psychophysiology: Human
  behavior and physiological response .}}
\newblock {\emph{\JournalTitle{PsycNet}}}  (\bibinfo{year}{2001}).

\bibitem{hoffman2013adrenaline}
\bibinfo{author}{Hoffman, B.~B.}
\newblock \emph{\bibinfo{title}{Adrenaline}} (\bibinfo{publisher}{Harvard
  University Press}, \bibinfo{year}{2013}).

\bibitem{awada2024stress}
\bibinfo{author}{Awada, M.}, \bibinfo{author}{Becerik~Gerber, B.},
  \bibinfo{author}{Lucas, G.~M.} \& \bibinfo{author}{Roll, S.~C.}
\newblock \bibinfo{journal}{\bibinfo{title}{Stress appraisal in the workplace
  and its associations with productivity and mood: Insights from a multimodal
  machine learning analysis}}.
\newblock {\emph{\JournalTitle{Plos one}}} \textbf{\bibinfo{volume}{19}},
  \bibinfo{pages}{e0296468} (\bibinfo{year}{2024}).

\bibitem{li2016eustress}
\bibinfo{author}{Li, C.-T.}, \bibinfo{author}{Cao, J.} \& \bibinfo{author}{Li,
  T.~M.}
\newblock \bibinfo{title}{Eustress or distress: An empirical study of perceived
  stress in everyday college life}.
\newblock In \emph{\bibinfo{booktitle}{Proceedings of the 2016 ACM
  International Joint Conference on Pervasive and Ubiquitous Computing:
  Adjunct}}, \bibinfo{pages}{1209--1217} (\bibinfo{year}{2016}).

\bibitem{qian2014does}
\bibinfo{author}{Qian, X.~L.}, \bibinfo{author}{Yarnal, C.~M.} \&
  \bibinfo{author}{Almeida, D.~M.}
\newblock \bibinfo{journal}{\bibinfo{title}{Does leisure time moderate or
  mediate the effect of daily stress on positve affect? an examination using
  eight-day diary data}}.
\newblock {\emph{\JournalTitle{Journal of leisure research}}}
  \textbf{\bibinfo{volume}{46}}, \bibinfo{pages}{106--124}
  (\bibinfo{year}{2014}).

\bibitem{hill2018sense}
\bibinfo{author}{Hill, P.~L.}, \bibinfo{author}{Sin, N.~L.},
  \bibinfo{author}{Turiano, N.~A.}, \bibinfo{author}{Burrow, A.~L.} \&
  \bibinfo{author}{Almeida, D.~M.}
\newblock \bibinfo{journal}{\bibinfo{title}{Sense of purpose moderates the
  associations between daily stressors and daily well-being}}.
\newblock {\emph{\JournalTitle{Annals of Behavioral Medicine}}}
  \textbf{\bibinfo{volume}{52}}, \bibinfo{pages}{724--729}
  (\bibinfo{year}{2018}).

\bibitem{jin2020impact}
\bibinfo{author}{Jin, M.~J.} \emph{et~al.}
\newblock \bibinfo{journal}{\bibinfo{title}{The impact of emotional exhaustion
  on psychological factors in workers with secondary traumatic experiences: A
  multi-group path analysis}}.
\newblock {\emph{\JournalTitle{Psychiatry Investigation}}}
  \textbf{\bibinfo{volume}{17}}, \bibinfo{pages}{1064} (\bibinfo{year}{2020}).

\bibitem{mcewen2006protective}
\bibinfo{author}{McEwen, B.~S.}
\newblock \bibinfo{journal}{\bibinfo{title}{Protective and damaging effects of
  stress mediators: central role of the brain}}.
\newblock {\emph{\JournalTitle{Dialogues in clinical neuroscience}}}
  \textbf{\bibinfo{volume}{8}}, \bibinfo{pages}{367--381}
  (\bibinfo{year}{2006}).

\bibitem{schneiderman2005stress}
\bibinfo{author}{Schneiderman, N.}, \bibinfo{author}{Ironson, G.} \&
  \bibinfo{author}{Siegel, S.~D.}
\newblock \bibinfo{journal}{\bibinfo{title}{Stress and health: psychological,
  behavioral, and biological determinants}}.
\newblock {\emph{\JournalTitle{Annu. Rev. Clin. Psychol.}}}
  \textbf{\bibinfo{volume}{1}}, \bibinfo{pages}{607--628}
  (\bibinfo{year}{2005}).

\bibitem{maslach1998multidimensional}
\bibinfo{author}{Maslach, C.}
\newblock \bibinfo{journal}{\bibinfo{title}{A multidimensional theory of
  burnout}}.
\newblock {\emph{\JournalTitle{Theories of organizational stress}}}
  \textbf{\bibinfo{volume}{68}}, \bibinfo{pages}{16} (\bibinfo{year}{1998}).

\bibitem{murphy2003usa}
\bibinfo{author}{Murphy, L.~R.} \& \bibinfo{author}{Sauter, S.~L.}
\newblock \bibinfo{journal}{\bibinfo{title}{The usa perspective: Current issues
  and trends in the management of work stress}}.
\newblock {\emph{\JournalTitle{Australian Psychologist}}}
  \textbf{\bibinfo{volume}{38}}, \bibinfo{pages}{151--157}
  (\bibinfo{year}{2003}).

\bibitem{schmidt2018introducing}
\bibinfo{author}{Schmidt, P.}, \bibinfo{author}{Reiss, A.},
  \bibinfo{author}{Duerichen, R.}, \bibinfo{author}{Marberger, C.} \&
  \bibinfo{author}{Van~Laerhoven, K.}
\newblock \bibinfo{title}{Introducing wesad, a multimodal dataset for wearable
  stress and affect detection}.
\newblock In \emph{\bibinfo{booktitle}{Proceedings of the 20th ACM
  international conference on multimodal interaction}},
  \bibinfo{pages}{400--408} (\bibinfo{year}{2018}).

\bibitem{hosseini2021multi}
\bibinfo{author}{Hosseini, S.} \emph{et~al.}
\newblock \bibinfo{journal}{\bibinfo{title}{A multi-modal sensor dataset for
  continuous stress detection of nurses in a hospital}}.
\newblock {\emph{\JournalTitle{arXiv preprint arXiv:2108.07689}}}
  (\bibinfo{year}{2021}).

\bibitem{sohrab2020facial}
\bibinfo{author}{Sohrab, F.}, \bibinfo{author}{Raitoharju, J.} \&
  \bibinfo{author}{Gabbouj, M.}
\newblock \bibinfo{title}{Facial expression based satisfaction index for
  empathic buildings}.
\newblock In \emph{\bibinfo{booktitle}{Adjunct Proceedings of the 2020 ACM
  International Joint Conference on Pervasive and Ubiquitous Computing and
  Proceedings of the 2020 ACM International Symposium on Wearable Computers}},
  \bibinfo{pages}{704--707} (\bibinfo{year}{2020}).

\bibitem{sioni2015stress}
\bibinfo{author}{Sioni, R.} \& \bibinfo{author}{Chittaro, L.}
\newblock \bibinfo{journal}{\bibinfo{title}{Stress detection using
  physiological sensors}}.
\newblock {\emph{\JournalTitle{Computer}}} \textbf{\bibinfo{volume}{48}},
  \bibinfo{pages}{26--33} (\bibinfo{year}{2015}).

\bibitem{thapliyal2017stress}
\bibinfo{author}{Thapliyal, H.}, \bibinfo{author}{Khalus, V.} \&
  \bibinfo{author}{Labrado, C.}
\newblock \bibinfo{journal}{\bibinfo{title}{Stress detection and management: A
  survey of wearable smart health devices}}.
\newblock {\emph{\JournalTitle{IEEE Consumer Electronics Magazine}}}
  \textbf{\bibinfo{volume}{6}}, \bibinfo{pages}{64--69} (\bibinfo{year}{2017}).

\bibitem{carneiro2017new}
\bibinfo{author}{Carneiro, D.}, \bibinfo{author}{Novais, P.},
  \bibinfo{author}{Augusto, J.~C.} \& \bibinfo{author}{Payne, N.}
\newblock \bibinfo{journal}{\bibinfo{title}{New methods for stress assessment
  and monitoring at the workplace}}.
\newblock {\emph{\JournalTitle{IEEE Transactions on Affective Computing}}}
  \textbf{\bibinfo{volume}{10}}, \bibinfo{pages}{237--254}
  (\bibinfo{year}{2017}).

\bibitem{alberdi2016towards}
\bibinfo{author}{Alberdi, A.}, \bibinfo{author}{Aztiria, A.} \&
  \bibinfo{author}{Basarab, A.}
\newblock \bibinfo{journal}{\bibinfo{title}{Towards an automatic early stress
  recognition system for office environments based on multimodal measurements:
  A review}}.
\newblock {\emph{\JournalTitle{Journal of biomedical informatics}}}
  \textbf{\bibinfo{volume}{59}}, \bibinfo{pages}{49--75}
  (\bibinfo{year}{2016}).

\bibitem{fukazawa2019predicting}
\bibinfo{author}{Fukazawa, Y.} \emph{et~al.}
\newblock \bibinfo{journal}{\bibinfo{title}{Predicting anxiety state using
  smartphone-based passive sensing}}.
\newblock {\emph{\JournalTitle{Journal of biomedical informatics}}}
  \textbf{\bibinfo{volume}{93}}, \bibinfo{pages}{103151}
  (\bibinfo{year}{2019}).

\bibitem{panicker2019survey}
\bibinfo{author}{Panicker, S.~S.} \& \bibinfo{author}{Gayathri, P.}
\newblock \bibinfo{journal}{\bibinfo{title}{A survey of machine learning
  techniques in physiology based mental stress detection systems}}.
\newblock {\emph{\JournalTitle{Biocybernetics and Biomedical Engineering}}}
  \textbf{\bibinfo{volume}{39}}, \bibinfo{pages}{444--469}
  (\bibinfo{year}{2019}).

\bibitem{zhang2020emotion}
\bibinfo{author}{Zhang, X.} \emph{et~al.}
\newblock \bibinfo{journal}{\bibinfo{title}{Emotion recognition from multimodal
  physiological signals using a regularized deep fusion of kernel machine}}.
\newblock {\emph{\JournalTitle{IEEE transactions on cybernetics}}}
  \textbf{\bibinfo{volume}{51}}, \bibinfo{pages}{4386--4399}
  (\bibinfo{year}{2020}).

\bibitem{kurniawan2013stress}
\bibinfo{author}{Kurniawan, H.}, \bibinfo{author}{Maslov, A.~V.} \&
  \bibinfo{author}{Pechenizkiy, M.}
\newblock \bibinfo{title}{Stress detection from speech and galvanic skin
  response signals}.
\newblock In \emph{\bibinfo{booktitle}{Proceedings of the 26th IEEE
  International Symposium on Computer-Based Medical Systems}},
  \bibinfo{pages}{209--214} (\bibinfo{organization}{IEEE},
  \bibinfo{year}{2013}).

\bibitem{pavlidis2005system}
\bibinfo{author}{Pavlidis, I.}
\newblock \bibinfo{title}{System and method using thermal image analysis for
  polygraph testing} (\bibinfo{year}{2005}).
\newblock \bibinfo{note}{US Patent 6,854,879}.

\bibitem{haouij2018affectiveroad}
\bibinfo{author}{Haouij, N.~E.}, \bibinfo{author}{Poggi, J.-M.},
  \bibinfo{author}{Sevestre-Ghalila, S.}, \bibinfo{author}{Ghozi, R.} \&
  \bibinfo{author}{Ja{\"\i}dane, M.}
\newblock \bibinfo{title}{Affectiveroad system and database to assess driver's
  attention}.
\newblock In \emph{\bibinfo{booktitle}{Proceedings of the 33rd Annual ACM
  Symposium on Applied Computing}}, \bibinfo{pages}{800--803}
  (\bibinfo{year}{2018}).

\bibitem{kirschbaum1993trier}
\bibinfo{author}{Kirschbaum, C.}, \bibinfo{author}{Pirke, K.} \&
  \bibinfo{author}{Hellhammer, D.}
\newblock \bibinfo{journal}{\bibinfo{title}{The 'trier social stress test'—a
  tool for investigating psychobiological stress responses in a laboratory
  setting}}.
\newblock {\emph{\JournalTitle{Neuropsychobiology}}}
  \textbf{\bibinfo{volume}{28}}, \bibinfo{pages}{76--81}
  (\bibinfo{year}{1993}).

\bibitem{chen2021introducing}
\bibinfo{author}{Chen, W.}, \bibinfo{author}{Zheng, S.} \&
  \bibinfo{author}{Sun, X.}
\newblock \bibinfo{title}{Introducing mdpsd, a multimodal dataset for
  psychological stress detection}.
\newblock In \emph{\bibinfo{booktitle}{Big Data: 8th CCF Conference, BigData
  2020, Chongqing, China, October 22--24, 2020, Revised Selected Papers}}, vol.
  \bibinfo{volume}{1320}, \bibinfo{pages}{59} (\bibinfo{organization}{Springer
  Nature}, \bibinfo{year}{2021}).

\bibitem{birkett2011trier}
\bibinfo{author}{Birkett, M.~A.}
\newblock \bibinfo{journal}{\bibinfo{title}{The trier social stress test
  protocol for inducing psychological stress}}.
\newblock {\emph{\JournalTitle{JoVE (Journal of Visualized Experiments)}}}
  \bibinfo{pages}{e3238} (\bibinfo{year}{2011}).

\bibitem{scarpina2017stroop}
\bibinfo{author}{Scarpina, F.} \& \bibinfo{author}{Tagini, S.}
\newblock \bibinfo{journal}{\bibinfo{title}{The stroop color and word test}}.
\newblock {\emph{\JournalTitle{Frontiers in Psychology}}}
  \textbf{\bibinfo{volume}{8}}, \bibinfo{pages}{557},
  \doiprefix\url{10.3389/fpsyg.2017.00557} (\bibinfo{year}{2017}).

\bibitem{mundnich2020tiles}
\bibinfo{author}{Mundnich, K.} \emph{et~al.}
\newblock \bibinfo{journal}{\bibinfo{title}{Tiles-2018, a longitudinal
  physiologic and behavioral data set of hospital workers}}.
\newblock {\emph{\JournalTitle{Scientific Data}}} \textbf{\bibinfo{volume}{7}},
  \bibinfo{pages}{1--26} (\bibinfo{year}{2020}).

\bibitem{koldijk2014swell}
\bibinfo{author}{Koldijk, S.}, \bibinfo{author}{Sappelli, M.},
  \bibinfo{author}{Verberne, S.}, \bibinfo{author}{Neerincx, M.~A.} \&
  \bibinfo{author}{Kraaij, W.}
\newblock \bibinfo{title}{The swell knowledge work dataset for stress and user
  modeling research}.
\newblock In \emph{\bibinfo{booktitle}{Proceedings of the 16th international
  conference on multimodal interaction}}, \bibinfo{pages}{291--298}
  (\bibinfo{year}{2014}).

\bibitem{zaman2019stress}
\bibinfo{author}{Zaman, S.} \emph{et~al.}
\newblock \bibinfo{journal}{\bibinfo{title}{Stress and productivity patterns of
  interrupted, synergistic, and antagonistic office activities}}.
\newblock {\emph{\JournalTitle{Scientific data}}} \textbf{\bibinfo{volume}{6}},
  \bibinfo{pages}{264} (\bibinfo{year}{2019}).

\bibitem{vanvoorhis2007understanding}
\bibinfo{author}{VanVoorhis, C. R.~W.} \& \bibinfo{author}{Morgan, B.~L.}
\newblock \bibinfo{journal}{\bibinfo{title}{Understanding power and rules of
  thumb for determining sample sizes}}.
\newblock {\emph{\JournalTitle{Tutor Quant Methods Psychol}}}
  \textbf{\bibinfo{volume}{3}}, \bibinfo{pages}{43--50} (\bibinfo{year}{2007}).

\bibitem{viola2001rapid}
\bibinfo{author}{Viola, P.} \& \bibinfo{author}{Jones, M.}
\newblock \bibinfo{title}{Rapid object detection using a boosted cascade of
  simple features}.
\newblock In \emph{\bibinfo{booktitle}{Proceedings of the 2001 IEEE computer
  society conference on computer vision and pattern recognition. CVPR 2001}},
  vol.~\bibinfo{volume}{1}, \bibinfo{pages}{I--I}
  (\bibinfo{organization}{Ieee}, \bibinfo{year}{2001}).

\bibitem{bradski2000opencv}
\bibinfo{author}{Bradski, G.}
\newblock \bibinfo{journal}{\bibinfo{title}{The opencv library.}}
\newblock {\emph{\JournalTitle{Dr. Dobb's Journal: Software Tools for the
  Professional Programmer}}} \textbf{\bibinfo{volume}{25}},
  \bibinfo{pages}{120--123} (\bibinfo{year}{2000}).

\bibitem{padilla2012evaluation}
\bibinfo{author}{Padilla, R.}, \bibinfo{author}{Costa~Filho, C.} \&
  \bibinfo{author}{Costa, M.}
\newblock \bibinfo{journal}{\bibinfo{title}{Evaluation of haar cascade
  classifiers designed for face detection}}.
\newblock {\emph{\JournalTitle{World Academy of Science, Engineering and
  Technology}}} \textbf{\bibinfo{volume}{64}}, \bibinfo{pages}{362--365}
  (\bibinfo{year}{2012}).

\bibitem{arriaga2017real}
\bibinfo{author}{Arriaga, O.}, \bibinfo{author}{Valdenegro-Toro, M.} \&
  \bibinfo{author}{Pl{\"o}ger, P.}
\newblock \bibinfo{journal}{\bibinfo{title}{Real-time convolutional neural
  networks for emotion and gender classification}}.
\newblock {\emph{\JournalTitle{arXiv preprint arXiv:1710.07557}}}
  (\bibinfo{year}{2017}).

\bibitem{goodfellow2013challenges}
\bibinfo{author}{Goodfellow, I.~J.} \emph{et~al.}
\newblock \bibinfo{title}{Challenges in representation learning: A report on
  three machine learning contests}.
\newblock In \emph{\bibinfo{booktitle}{International conference on neural
  information processing}}, \bibinfo{pages}{117--124}
  (\bibinfo{organization}{Springer}, \bibinfo{year}{2013}).

\bibitem{mohanty2019design}
\bibinfo{author}{Mohanty, S.}, \bibinfo{author}{Hegde, S.~V.},
  \bibinfo{author}{Prasad, S.} \& \bibinfo{author}{Manikandan, J.}
\newblock \bibinfo{title}{Design of real-time drowsiness detection system using
  dlib}.
\newblock In \emph{\bibinfo{booktitle}{2019 IEEE International WIE Conference
  on Electrical and Computer Engineering (WIECON-ECE)}}, \bibinfo{pages}{1--4}
  (\bibinfo{organization}{IEEE}, \bibinfo{year}{2019}).

\bibitem{irtija2018fatigue}
\bibinfo{author}{Irtija, N.}, \bibinfo{author}{Sami, M.} \&
  \bibinfo{author}{Ahad, M. A.~R.}
\newblock \bibinfo{title}{Fatigue detection using facial landmarks}.
\newblock In \emph{\bibinfo{booktitle}{International Symposium on Affective
  Science and Engineering ISASE2018}}, \bibinfo{pages}{1--6}
  (\bibinfo{organization}{Japan Society of Kansei Engineering},
  \bibinfo{year}{2018}).

\bibitem{heidari202progressive}
\bibinfo{author}{Heidari, N.} \& \bibinfo{author}{Iosifidis, A.}
\newblock \bibinfo{title}{Progressive spatio-temporal bilinear network with
  monte carlo dropout for landmark-based facial expression recognition with
  uncertainty estimation}.
\newblock In \emph{\bibinfo{booktitle}{IEEE International Workshop on
  Multimedia Signal Processing}} (\bibinfo{organization}{IEEE},
  \bibinfo{year}{2021}).

\bibitem{kumar2021micro}
\bibinfo{author}{Kumar, A. J.~R.} \& \bibinfo{author}{Bhanu, B.}
\newblock \bibinfo{title}{Micro-expression classification based on landmark
  relations with graph attention convolutional network}.
\newblock In \emph{\bibinfo{booktitle}{Proceedings of the IEEE/CVF Conference
  on Computer Vision and Pattern Recognition}}, \bibinfo{pages}{1511--1520}
  (\bibinfo{year}{2021}).

\bibitem{king2009dlib}
\bibinfo{author}{King, D.~E.}
\newblock \bibinfo{journal}{\bibinfo{title}{Dlib-ml: A machine learning
  toolkit}}.
\newblock {\emph{\JournalTitle{The Journal of Machine Learning Research}}}
  \textbf{\bibinfo{volume}{10}}, \bibinfo{pages}{1755--1758}
  (\bibinfo{year}{2009}).

\bibitem{giannakakis2020automatic}
\bibinfo{author}{Giannakakis, G.}, \bibinfo{author}{Koujan, M.~R.},
  \bibinfo{author}{Roussos, A.} \& \bibinfo{author}{Marias, K.}
\newblock \bibinfo{title}{Automatic stress detection evaluating models of
  facial action units}.
\newblock In \emph{\bibinfo{booktitle}{2020 15th IEEE international conference
  on automatic face and gesture recognition (FG 2020)}},
  \bibinfo{pages}{728--733} (\bibinfo{organization}{IEEE},
  \bibinfo{year}{2020}).

\bibitem{PredictingDASS2019}
\bibinfo{author}{Authors, A.}
\newblock \bibinfo{journal}{\bibinfo{title}{Predicting depression, anxiety, and
  stress levels from videos using the facial action coding system}}.
\newblock {\emph{\JournalTitle{Sensors}}} \textbf{\bibinfo{volume}{19}},
  \bibinfo{pages}{3693}, \doiprefix\url{10.3390/s19173693}
  (\bibinfo{year}{2019}).

\bibitem{AutomaticStressAU2020}
\bibinfo{author}{Authors, A.}
\newblock \bibinfo{title}{Automatic stress detection evaluating models of
  facial action units}.
\newblock In \emph{\bibinfo{booktitle}{2020 International Conference on
  Automatic Face and Gesture Recognition (FG)}} (\bibinfo{year}{2020}).

\bibitem{AutomaticStressFACS2021}
\bibinfo{author}{Authors, A.}
\newblock \bibinfo{journal}{\bibinfo{title}{Automatic stress analysis from
  facial videos based on deep facial action units recognition}}.
\newblock {\emph{\JournalTitle{Pattern Analysis and Applications}}}
  \doiprefix\url{10.1007/s10044-021-01012-9} (\bibinfo{year}{2021}).

\bibitem{Majid2022}
\bibinfo{author}{CPHSLab}.
\newblock \bibinfo{title}{Empathicschool}.
\newblock
  \bibinfo{howpublished}{\url{https://github.com/CPHSLab/EmpathicSchool}}
  (\bibinfo{year}{2022}).

\bibitem{fox2007resting}
\bibinfo{author}{Fox, K.} \emph{et~al.}
\newblock \bibinfo{journal}{\bibinfo{title}{Resting heart rate in
  cardiovascular disease}}.
\newblock {\emph{\JournalTitle{Journal of the American College of Cardiology}}}
  \textbf{\bibinfo{volume}{50}}, \bibinfo{pages}{823--830}
  (\bibinfo{year}{2007}).

\bibitem{shi2017differences}
\bibinfo{author}{Shi, H.} \emph{et~al.}
\newblock \bibinfo{journal}{\bibinfo{title}{Differences of heart rate
  variability between happiness and sadness emotion states: a pilot study}}.
\newblock {\emph{\JournalTitle{Journal of Medical and Biological Engineering}}}
  \textbf{\bibinfo{volume}{37}}, \bibinfo{pages}{527--539}
  (\bibinfo{year}{2017}).

\bibitem{taelman2009influence}
\bibinfo{author}{Taelman, J.}, \bibinfo{author}{Vandeput, S.},
  \bibinfo{author}{Spaepen, A.} \& \bibinfo{author}{Van~Huffel, S.}
\newblock \bibinfo{title}{Influence of mental stress on heart rate and heart
  rate variability}.
\newblock In \emph{\bibinfo{booktitle}{4th European Conference of the
  International Federation for Medical and Biological Engineering}},
  vol.~\bibinfo{volume}{22} of \emph{\bibinfo{series}{IFMBE Proceedings}},
  \bibinfo{pages}{1366--1369}, \doiprefix\url{10.1007/978-3-540-89208-3\_324}
  (\bibinfo{year}{2009}).

\bibitem{Kim2018StressHRV}
\bibinfo{author}{Kim, H.-G.}, \bibinfo{author}{Cheon, E.-J.},
  \bibinfo{author}{Bai, D.-S.}, \bibinfo{author}{Lee, Y.-H.} \&
  \bibinfo{author}{Koo, B.-H.}
\newblock \bibinfo{journal}{\bibinfo{title}{Stress and heart rate variability:
  A meta-analysis and review of the literature}}.
\newblock {\emph{\JournalTitle{Psychiatry Investigation}}}
  \textbf{\bibinfo{volume}{15}}, \bibinfo{pages}{235--245},
  \doiprefix\url{10.30773/pi.2017.08.17} (\bibinfo{year}{2018}).

\bibitem{schubert2009effects}
\bibinfo{author}{Schubert, C.} \emph{et~al.}
\newblock \bibinfo{journal}{\bibinfo{title}{Effects of stress on heart rate
  complexity—a comparison between short-term and chronic stress}}.
\newblock {\emph{\JournalTitle{Biological psychology}}}
  \textbf{\bibinfo{volume}{80}}, \bibinfo{pages}{325--332}
  (\bibinfo{year}{2009}).

\bibitem{tanda2021simplified}
\bibinfo{author}{Tanda, G.}
\newblock \bibinfo{journal}{\bibinfo{title}{A simplified approach to describe
  the mean skin temperature variations during prolonged running exercise}}.
\newblock {\emph{\JournalTitle{Journal of Thermal Biology}}}
  \textbf{\bibinfo{volume}{99}}, \bibinfo{pages}{103005}
  (\bibinfo{year}{2021}).

\bibitem{bierman1936temperature}
\bibinfo{author}{Bierman, W.}
\newblock \bibinfo{journal}{\bibinfo{title}{The temperature of the skin
  surface}}.
\newblock {\emph{\JournalTitle{Journal of the American Medical Association}}}
  \textbf{\bibinfo{volume}{106}}, \bibinfo{pages}{1158--1162}
  (\bibinfo{year}{1936}).

\bibitem{choi2013skin}
\bibinfo{author}{Choi, J.} \& \bibinfo{author}{Yeom, S.}
\newblock \bibinfo{journal}{\bibinfo{title}{Skin temperature changes during
  physical activities and passive heating}}.
\newblock {\emph{\JournalTitle{Journal of Physical Therapy Science}}}
  \textbf{\bibinfo{volume}{25}}, \bibinfo{pages}{1029--1031},
  \doiprefix\url{10.1589/jpts.25.1029} (\bibinfo{year}{2013}).

\bibitem{stadler2018electrodermal}
\bibinfo{author}{Stadler, R.}, \bibinfo{author}{Jepson, A.~S.} \&
  \bibinfo{author}{Wood, E.~H.}
\newblock \bibinfo{journal}{\bibinfo{title}{Electrodermal activity measurement
  within a qualitative methodology: Exploring emotion in leisure experiences}}.
\newblock {\emph{\JournalTitle{International Journal of Contemporary
  Hospitality Management}}} \textbf{\bibinfo{volume}{30}},
  \bibinfo{pages}{3363--3385} (\bibinfo{year}{2018}).

\bibitem{Boucsein2012Electrodermal}
\bibinfo{author}{Boucsein, W.}
\newblock \emph{\bibinfo{title}{Electrodermal Activity}}
  (\bibinfo{publisher}{Springer}, \bibinfo{address}{New York, NY},
  \bibinfo{year}{2012}), \bibinfo{edition}{2nd} edn.

\bibitem{bayat2014study}
\bibinfo{author}{Bayat, A.}, \bibinfo{author}{Pomplun, M.} \&
  \bibinfo{author}{Tran, D.~A.}
\newblock \bibinfo{journal}{\bibinfo{title}{A study on human activity
  recognition using accelerometer data from smartphones}}.
\newblock {\emph{\JournalTitle{Procedia Computer Science}}}
  \textbf{\bibinfo{volume}{34}}, \bibinfo{pages}{450--457}
  (\bibinfo{year}{2014}).

\bibitem{cudejko2022validity}
\bibinfo{author}{Cudejko, T.}, \bibinfo{author}{Button, K.} \&
  \bibinfo{author}{Al-Amri, M.}
\newblock \bibinfo{journal}{\bibinfo{title}{Validity and reliability of
  accelerations and orientations measured using wearable sensors during
  functional activities}}.
\newblock {\emph{\JournalTitle{Scientific reports}}}
  \textbf{\bibinfo{volume}{12}}, \bibinfo{pages}{14619} (\bibinfo{year}{2022}).

\bibitem{abay2018photoplethysmography}
\bibinfo{author}{Abay, T.~Y.} \& \bibinfo{author}{Kyriacou, P.}
\newblock \bibinfo{journal}{\bibinfo{title}{Photoplethysmography for blood
  volumes and oxygenation changes during intermittent vascular occlusions}}.
\newblock {\emph{\JournalTitle{Journal of clinical monitoring and computing}}}
  \textbf{\bibinfo{volume}{32}}, \bibinfo{pages}{447--455}
  (\bibinfo{year}{2018}).

\bibitem{knight2001relaxing}
\bibinfo{author}{Knight, D.} \& \bibinfo{author}{Rickard, N.}
\newblock \bibinfo{journal}{\bibinfo{title}{Relaxing effects of reading and
  listening to music after a brief stressor}}.
\newblock {\emph{\JournalTitle{Australian Journal of Psychology}}}
  \textbf{\bibinfo{volume}{53}}, \bibinfo{pages}{112--117}
  (\bibinfo{year}{2001}).

\bibitem{ferrer2014playing}
\bibinfo{author}{Ferrer, A.}
\newblock \bibinfo{journal}{\bibinfo{title}{Playing music reduces salivary
  cortisol levels and improves mood after acute stress}}.
\newblock {\emph{\JournalTitle{Medical Science Monitor}}}
  \textbf{\bibinfo{volume}{20}}, \bibinfo{pages}{721--727}
  (\bibinfo{year}{2014}).

\bibitem{allen2017trier}
\bibinfo{author}{Allen, A.~P.} \emph{et~al.}
\newblock \bibinfo{journal}{\bibinfo{title}{The trier social stress test:
  principles and practice}}.
\newblock {\emph{\JournalTitle{Neurobiology of stress}}}
  \textbf{\bibinfo{volume}{6}}, \bibinfo{pages}{113--126}
  (\bibinfo{year}{2017}).

\bibitem{Allen2017tsstReview}
\bibinfo{author}{Allen, A. P. e.~a.}
\newblock \bibinfo{journal}{\bibinfo{title}{The trier social stress test:
  Principles and practice}}.
\newblock {\emph{\JournalTitle{Neurobiology of Stress}}}
  \textbf{\bibinfo{volume}{6}}, \bibinfo{pages}{113--126}
  (\bibinfo{year}{2017}).

\bibitem{callinan1996stroopStress}
\bibinfo{author}{Renaud, P.} \& \bibinfo{author}{Blondin, J.-P.}
\newblock \bibinfo{journal}{\bibinfo{title}{The stress of stroop performance:
  Physiological and emotional responses to color–word interference, task
  pacing, and pacing speed}}.
\newblock {\emph{\JournalTitle{International Journal of Psychophysiology}}}
  \textbf{\bibinfo{volume}{27}}, \bibinfo{pages}{87--97},
  \doiprefix\url{10.1016/S0167-8760(97)00049-4} (\bibinfo{year}{1997}).

\bibitem{laborde2021breathing}
\bibinfo{author}{Laborde, S.}
\newblock \bibinfo{journal}{\bibinfo{title}{Slow‐paced breathing for stress
  management}}.
\newblock {\emph{\JournalTitle{Psychophysiology}}}
  \textbf{\bibinfo{volume}{58}}, \doiprefix\url{10.1111/psyp.13770}
  (\bibinfo{year}{2021}).

\bibitem{pasatReview2018}
\bibinfo{author}{Tombaugh, T.~N.}
\newblock \bibinfo{title}{The paced auditory serial addition test (pasat): A
  review and update}.
\newblock In \emph{\bibinfo{booktitle}{Clinical Tests of Working Memory}}
  (\bibinfo{publisher}{Springer}, \bibinfo{year}{2018}).

\bibitem{hart2006nasa}
\bibinfo{author}{Hart, S.~G.}
\newblock \bibinfo{title}{Nasa-task load index (nasa-tlx); 20 years later}.
\newblock In \emph{\bibinfo{booktitle}{Proceedings of the human factors and
  ergonomics society annual meeting}}, vol.~\bibinfo{volume}{50},
  \bibinfo{pages}{904--908} (\bibinfo{organization}{Sage publications Sage CA:
  Los Angeles, CA}, \bibinfo{year}{2006}).

\bibitem{hart1988development}
\bibinfo{author}{Hart, S.~G.} \& \bibinfo{author}{Staveland, L.~E.}
\newblock \bibinfo{title}{Development of nasa-tlx (task load index): Results of
  empirical and theoretical research}.
\newblock In \emph{\bibinfo{booktitle}{Advances in psychology}},
  vol.~\bibinfo{volume}{52}, \bibinfo{pages}{139--183}
  (\bibinfo{publisher}{Elsevier}, \bibinfo{year}{1988}).

\bibitem{tavakol2011making}
\bibinfo{author}{Tavakol, M.} \& \bibinfo{author}{Dennick, R.}
\newblock \bibinfo{journal}{\bibinfo{title}{Making sense of cronbach's alpha}}.
\newblock {\emph{\JournalTitle{International journal of medical education}}}
  \textbf{\bibinfo{volume}{2}}, \bibinfo{pages}{53} (\bibinfo{year}{2011}).

\bibitem{ZenodoDataset}
\bibinfo{author}{Hosseini, M.} \emph{et~al.}
\newblock \bibinfo{title}{Empathicschool: A multimodal dataset for real-time
  facial expressions and physiological data analysis under different stress
  conditions}, \doiprefix\url{10.5281/zenodo.15556502} (\bibinfo{year}{2023}).

\bibitem{hellevik2016extreme}
\bibinfo{author}{Hellevik, O.}
\newblock \bibinfo{journal}{\bibinfo{title}{Extreme nonresponse and response
  bias: A “worst case” analysis}}.
\newblock {\emph{\JournalTitle{Quality \& Quantity}}}
  \textbf{\bibinfo{volume}{50}}, \bibinfo{pages}{1969--1991}
  (\bibinfo{year}{2016}).

\bibitem{hjortsjo1969man}
\bibinfo{author}{Hjortsjo, C.-H.}
\newblock \bibinfo{journal}{\bibinfo{title}{Man’s face and mimic language.}}
\newblock {\emph{\JournalTitle{Studen litteratur.}}}  (\bibinfo{year}{1969}).

\bibitem{nicolle2012robust}
\bibinfo{author}{Nicolle, J.}, \bibinfo{author}{Rapp, V.},
  \bibinfo{author}{Bailly, K.}, \bibinfo{author}{Prevost, L.} \&
  \bibinfo{author}{Chetouani, M.}
\newblock \bibinfo{title}{Robust continuous prediction of human emotions using
  multiscale dynamic cues}.
\newblock In \emph{\bibinfo{booktitle}{Proceedings of the 14th ACM
  international conference on Multimodal interaction}},
  \bibinfo{pages}{501--508} (\bibinfo{year}{2012}).

\bibitem{schafer2011savitzky}
\bibinfo{author}{Schafer, R.~W.}
\newblock \bibinfo{journal}{\bibinfo{title}{What is a savitzky-golay
  filter?[lecture notes]}}.
\newblock {\emph{\JournalTitle{IEEE Signal processing magazine}}}
  \textbf{\bibinfo{volume}{28}}, \bibinfo{pages}{111--117}
  (\bibinfo{year}{2011}).

\bibitem{etkin2011emotional}
\bibinfo{author}{Etkin, A.}, \bibinfo{author}{Egner, T.} \&
  \bibinfo{author}{Kalisch, R.}
\newblock \bibinfo{journal}{\bibinfo{title}{Emotional processing in anterior
  cingulate and medial prefrontal cortex}}.
\newblock {\emph{\JournalTitle{Trends in cognitive sciences}}}
  \textbf{\bibinfo{volume}{15}}, \bibinfo{pages}{85--93}
  (\bibinfo{year}{2011}).

\bibitem{wang2015video}
\bibinfo{author}{Wang, S.} \& \bibinfo{author}{Ji, Q.}
\newblock \bibinfo{journal}{\bibinfo{title}{Video affective content analysis: A
  survey of state-of-the-art methods}}.
\newblock {\emph{\JournalTitle{IEEE Transactions on Affective Computing}}}
  \textbf{\bibinfo{volume}{6}}, \bibinfo{pages}{410--430}
  (\bibinfo{year}{2015}).

\bibitem{gjoreski2017continuous}
\bibinfo{author}{Gjoreski, M.}, \bibinfo{author}{Gjoreski, H.} \&
  \bibinfo{author}{Luštrek, M.}
\newblock \bibinfo{title}{Continuous stress detection using wearable sensors in
  real life: Algorithm selection and imu validation}.
\newblock In \emph{\bibinfo{booktitle}{Proceedings of the 2017 ACM
  International Joint Conference on Pervasive and Ubiquitous Computing}},
  UbiComp '17, \bibinfo{pages}{1185--1193} (\bibinfo{publisher}{ACM},
  \bibinfo{address}{New York, NY, USA}, \bibinfo{year}{2017}).

\bibitem{hernandez2011using}
\bibinfo{author}{Hern{\'a}ndez, J.~C.} \& \bibinfo{author}{Picard, R.~W.}
\newblock \bibinfo{title}{Using wearable sensors to detect stress in the wild}.
\newblock In \emph{\bibinfo{booktitle}{Proceedings of the 2011 IEEE
  International Conference on Affective Computing and Intelligent
  Interaction}}, ACII '11, \bibinfo{pages}{602--608}
  (\bibinfo{publisher}{IEEE}, \bibinfo{address}{Bethesda, MD, USA},
  \bibinfo{year}{2011}).

\bibitem{kim2018stress}
\bibinfo{author}{Kim, J.}, \bibinfo{author}{Garcia, E.~P.} \&
  \bibinfo{author}{Picard, R.~W.}
\newblock \bibinfo{journal}{\bibinfo{title}{Stress classification from wearable
  sensor data: Impact of feature selection and class imbalance}}.
\newblock {\emph{\JournalTitle{IEEE Journal of Biomedical and Health
  Informatics}}} \textbf{\bibinfo{volume}{22}}, \bibinfo{pages}{888--900},
  \doiprefix\url{10.1109/JBHI.2017.2654359} (\bibinfo{year}{2018}).

\bibitem{benedek2010continuous}
\bibinfo{author}{Benedek, M.} \& \bibinfo{author}{Kaernbach, C.}
\newblock \bibinfo{journal}{\bibinfo{title}{A continuous measure of phasic
  electrodermal activity}}.
\newblock {\emph{\JournalTitle{Journal of neuroscience methods}}}
  \textbf{\bibinfo{volume}{190}}, \bibinfo{pages}{80--91}
  (\bibinfo{year}{2010}).

\bibitem{mccraty2015heart}
\bibinfo{author}{McCraty, R.} \& \bibinfo{author}{Shaffer, F.}
\newblock \bibinfo{journal}{\bibinfo{title}{Heart rate variability: new
  perspectives on physiological mechanisms, assessment of self-regulatory
  capacity, and health risk}}.
\newblock {\emph{\JournalTitle{Global advances in health and medicine}}}
  \textbf{\bibinfo{volume}{4}}, \bibinfo{pages}{46--61} (\bibinfo{year}{2015}).

\bibitem{shaffer2017overview}
\bibinfo{author}{Shaffer, F.} \& \bibinfo{author}{Ginsberg, J.~P.}
\newblock \bibinfo{journal}{\bibinfo{title}{An overview of heart rate
  variability metrics and norms}}.
\newblock {\emph{\JournalTitle{Frontiers in Public Health}}}
  \textbf{\bibinfo{volume}{5}}, \bibinfo{pages}{258},
  \doiprefix\url{10.3389/fpubh.2017.00258} (\bibinfo{year}{2017}).

\bibitem{el2019random}
\bibinfo{author}{El~Haouij, N.}, \bibinfo{author}{Poggi, J.-M.},
  \bibinfo{author}{Ghozi, R.}, \bibinfo{author}{Sevestre-Ghalila, S.} \&
  \bibinfo{author}{Ja{\"\i}dane, M.}
\newblock \bibinfo{journal}{\bibinfo{title}{Random forest-based approach for
  physiological functional variable selection for driver’s stress level
  classification}}.
\newblock {\emph{\JournalTitle{Statistical Methods \& Applications}}}
  \textbf{\bibinfo{volume}{28}}, \bibinfo{pages}{157--185}
  (\bibinfo{year}{2019}).

\bibitem{chawla2002smote}
\bibinfo{author}{Chawla, N.~V.}, \bibinfo{author}{Bowyer, K.~W.},
  \bibinfo{author}{Hall, L.~O.} \& \bibinfo{author}{Kegelmeyer, W.~P.}
\newblock \bibinfo{journal}{\bibinfo{title}{Smote: Synthetic minority
  over-sampling technique}}.
\newblock {\emph{\JournalTitle{Journal of Artificial Intelligence Research}}}
  \textbf{\bibinfo{volume}{16}}, \bibinfo{pages}{321--357}
  (\bibinfo{year}{2002}).

\bibitem{salam2021effect}
\bibinfo{author}{Salam, M.~A.}, \bibinfo{author}{Azar, A.~T.},
  \bibinfo{author}{Elgendy, M.~S.} \& \bibinfo{author}{Fouad, K.~M.}
\newblock \bibinfo{journal}{\bibinfo{title}{The effect of different
  dimensionality reduction techniques on machine learning overfitting
  problem}}.
\newblock {\emph{\JournalTitle{Int. J. Adv. Comput. Sci. Appl}}}
  \textbf{\bibinfo{volume}{12}}, \bibinfo{pages}{641--655}
  (\bibinfo{year}{2021}).

\bibitem{benesty2009pearson}
\bibinfo{author}{Benesty, J.}, \bibinfo{author}{Chen, J.},
  \bibinfo{author}{Huang, Y.} \& \bibinfo{author}{Cohen, I.}
\newblock \bibinfo{title}{Pearson correlation coefficient}.
\newblock In \emph{\bibinfo{booktitle}{Noise reduction in speech processing}},
  \bibinfo{pages}{1--4} (\bibinfo{publisher}{Springer}, \bibinfo{year}{2009}).

\bibitem{wang2020heightened}
\bibinfo{author}{Wang, D.}, \bibinfo{author}{Schneider, S.},
  \bibinfo{author}{Schwartz, J.~E.} \& \bibinfo{author}{Stone, A.~A.}
\newblock \bibinfo{journal}{\bibinfo{title}{Heightened stress in employed
  individuals is linked to altered variability and inertia in emotions}}.
\newblock {\emph{\JournalTitle{Frontiers in psychology}}}
  \textbf{\bibinfo{volume}{11}}, \bibinfo{pages}{1152} (\bibinfo{year}{2020}).

\bibitem{McDuff2014RemoteStress}
\bibinfo{author}{McDuff, D.}, \bibinfo{author}{Gontarek, S.} \&
  \bibinfo{author}{Picard, R.}
\newblock \bibinfo{title}{Remote measurement of cognitive stress via heart rate
  variability}.
\newblock In \emph{\bibinfo{booktitle}{36th Annual International Conference of
  the IEEE Engineering in Medicine and Biology Society (EMBC 2014)}},
  \bibinfo{pages}{2957--2960}, \doiprefix\url{10.1109/EMBC.2014.6944243}
  (\bibinfo{year}{2014}).

\bibitem{vrijkotte2000effects}
\bibinfo{author}{Vrijkotte, T.~G.}, \bibinfo{author}{Van~Doornen, L.~J.} \&
  \bibinfo{author}{De~Geus, E.~J.}
\newblock \bibinfo{journal}{\bibinfo{title}{Effects of work stress on
  ambulatory blood pressure, heart rate, and heart rate variability}}.
\newblock {\emph{\JournalTitle{Hypertension}}} \textbf{\bibinfo{volume}{35}},
  \bibinfo{pages}{880--886} (\bibinfo{year}{2000}).

\bibitem{goodie2000validation}
\bibinfo{author}{Goodie, J.~L.}, \bibinfo{author}{Larkin, K.~T.} \&
  \bibinfo{author}{Schauss, S.}
\newblock \bibinfo{journal}{\bibinfo{title}{Validation of polar heart rate
  monitor for assessing heart rate during physical and mental stress.}}
\newblock {\emph{\JournalTitle{Journal of Psychophysiology}}}
  \textbf{\bibinfo{volume}{14}}, \bibinfo{pages}{159} (\bibinfo{year}{2000}).

\bibitem{chen2021pain}
\bibinfo{author}{Chen, J.}, \bibinfo{author}{Abbod, M.} \&
  \bibinfo{author}{Shieh, J.-S.}
\newblock \bibinfo{journal}{\bibinfo{title}{Pain and stress detection using
  wearable sensors and devices—a review}}.
\newblock {\emph{\JournalTitle{Sensors}}} \textbf{\bibinfo{volume}{21}},
  \bibinfo{pages}{1030} (\bibinfo{year}{2021}).

\bibitem{li2020stress}
\bibinfo{author}{Li, R.} \& \bibinfo{author}{Liu, Z.}
\newblock \bibinfo{journal}{\bibinfo{title}{Stress detection using deep neural
  networks}}.
\newblock {\emph{\JournalTitle{BMC Medical Informatics and Decision Making}}}
  \textbf{\bibinfo{volume}{20}}, \bibinfo{pages}{1--10} (\bibinfo{year}{2020}).

\bibitem{zhai2006stress}
\bibinfo{author}{Zhai, J.} \& \bibinfo{author}{Barreto, A.}
\newblock \bibinfo{title}{Stress detection in computer users based on digital
  signal processing of noninvasive physiological variables}.
\newblock In \emph{\bibinfo{booktitle}{2006 international conference of the
  IEEE engineering in medicine and biology society}},
  \bibinfo{pages}{1355--1358} (\bibinfo{organization}{IEEE},
  \bibinfo{year}{2006}).

\bibitem{yamakoshi2008feasibility}
\bibinfo{author}{Yamakoshi, T.} \emph{et~al.}
\newblock \bibinfo{title}{Feasibility study on driver's stress detection from
  differential skin temperature measurement}.
\newblock In \emph{\bibinfo{booktitle}{2008 30th annual international
  conference of the ieee engineering in medicine and biology society}},
  \bibinfo{pages}{1076--1079} (\bibinfo{organization}{IEEE},
  \bibinfo{year}{2008}).

\bibitem{tarrant1993effects}
\bibinfo{author}{Tarrant, M.~A.}, \bibinfo{author}{Manfredo, M.~J.},
  \bibinfo{author}{Bayley, P.~B.} \& \bibinfo{author}{Hess, R.}
\newblock \bibinfo{journal}{\bibinfo{title}{Effects of recall bias and
  nonresponse bias on self-report estimates of angling participation}}.
\newblock {\emph{\JournalTitle{North American Journal of Fisheries
  Management}}} \textbf{\bibinfo{volume}{13}}, \bibinfo{pages}{217--222}
  (\bibinfo{year}{1993}).

\bibitem{allen1999newcomer}
\bibinfo{author}{Allen, T.~D.}, \bibinfo{author}{McManus, S.~E.} \&
  \bibinfo{author}{Russell, J.~E.}
\newblock \bibinfo{journal}{\bibinfo{title}{Newcomer socialization and stress:
  Formal peer relationships as a source of support}}.
\newblock {\emph{\JournalTitle{Journal of Vocational Behavior}}}
  \textbf{\bibinfo{volume}{54}}, \bibinfo{pages}{453--470}
  (\bibinfo{year}{1999}).

\bibitem{szegedy2016rethinking}
\bibinfo{author}{Szegedy, C.}, \bibinfo{author}{Vanhoucke, V.},
  \bibinfo{author}{Ioffe, S.}, \bibinfo{author}{Shlens, J.} \&
  \bibinfo{author}{Wojna, Z.}
\newblock \bibinfo{title}{Rethinking the inception architecture for computer
  vision}.
\newblock In \emph{\bibinfo{booktitle}{Proceedings of the IEEE conference on
  computer vision and pattern recognition}}, \bibinfo{pages}{2818--2826}
  (\bibinfo{year}{2016}).

\end{thebibliography}

\end{document}